\documentclass[%
 aip,
 amsmath,amssymb,
 reprint,%
]{revtex4-1}

\usepackage{graphicx}
\usepackage{dcolumn}
\usepackage{bm}
\usepackage{mathrsfs}
\usepackage[utf8]{inputenc}
\usepackage[T1]{fontenc}
\usepackage{mathptmx}
\DeclareMathAlphabet{\mathcal}{OMS}{cmsy}{m}{n}
\usepackage{etoolbox}
\usepackage[caption=false, font=normalsize, labelfont=sf, textfont=sf,position=top,justification=raggedright,singlelinecheck=false]{subfig}
\makeatletter
\def\@email#1#2{%
 \endgroup
 \patchcmd{\titleblock@produce}
  {\frontmatter@RRAPformat}
  {\frontmatter@RRAPformat{\produce@RRAP{*#1\href{mailto:#2}{#2}}}\frontmatter@RRAPformat}
  {}{}
}%
\makeatother
\begin{document}

\preprint{AIP/123-QED}

\title[]{Theory of Electrically Detected Magnetic Resonance of Silicon Vacancy-Related Spin Pairs in Silicon Carbide}
\author{David A. Fehr}
 \email{david-fehr@uiowa.edu}
 \affiliation{Department of Physics and Astronomy,  University of Iowa, Iowa City, Iowa 52242, USA}%
\author{Corey J. Cochrane}%
 \affiliation{Jet Propulsion Laboratory, California Institute of Technology, Pasadena, California 91011, USA}%
\author{Stephen R. McMillan}
 \affiliation{Donostia International Physics Center, Donostia-San Sebastián, Spain}
\author{Nicholas J. Harmon}
 \affiliation{Department of Physics and Engineering Science, Coastal Carolina University, Conway, South Carolina 29528, USA}
\author{Patrick M. Lenahan}
 \affiliation{Department of Engineering Science and Mechanics, The Pennsylvania State University, University Park, Pennsylvania 16802, USA}
\author{Michael E. Flatt\'{e}}
 \affiliation{Department of Physics and Astronomy,  University of Iowa, Iowa City, Iowa 52242, USA}
 \altaffiliation{Department of Applied Physics, Eindhoven University of Technology, Eindhoven 5612 AZ, The Netherlands}

\date{\today}

\begin{abstract}
We present a quantitative theory for simulating the electrically detected magnetic resonance (EDMR)  of silicon vacancy-related spin pairs in silicon carbide using steady-state Lindblad master equations. In our theory, we consider V1a and V2a deep level silicon vacancies near the (0/-) charge state transition level in proximity to a previously identified nitrogen-related complex, the incomplete K-center, due to the hyperfine, spin structure, and Land\'e g factor of the shallow state. 
Our theory describes recent room temperature measurements attributed to V1a silicon vacancies, with reasonable extracted parameters for defect  spin coherence times and electrical transport rates. At lower temperatures we predict that the shallow level hyperfine structure may be spectrally resolvable. Finally, we predict the EDMR spectrum of V2a silicon vacancy-related spin pairs and predict that two-photon, double quantum transitions of the silicon vacancy's negative charge state can be electrically read-out for enhanced magnetic field sensing.
\end{abstract}

\maketitle
Silicon carbide (SiC) has emerged as an important material for power electronics beyond silicon due to its wide bandgap, enabling devices with higher operating temperatures and breakdown voltages, faster switching, and superior radiation hardness\cite{8257396}. SiC is also known to host electrically-active deep-level defects, both natively and resulting from irradiation, which limit device performance\cite{10.1063/1.2714285,10.1063/1.3530600}, and, in extreme cases, cause device failure\cite{PeartonS.J.2021RDiW}. On the other hand, some of these deep-level defects, namely silicon vacancies, have desirable quantum properties such as long coherence times\cite{WidmannMatthias2015Ccos}, optical\cite{TaharaKosuke2025Qswd,KrausH.2014Rqme} and electrical\cite{CochraneCoreyJ.2016Vmfs,GottschollAndreas2024Eter,10.1063/5.0154382,PhysRevLett.132.146902} readout modalities, and have great potential as quantum sensors of magnetic fields\cite{TaharaKosuke2025Qswd,CochraneCoreyJ.2016Vmfs,GottschollAndreas2024Eter,KrausH.2014Mfat,PhysRevApplied.19.044086,10.1063/5.0154382} and temperature\cite{KrausH.2014Mfat,PhysRevApplied.20.L031001}. Although defect-based quantum sensing schemes typically utilize optical readout for enhanced sensitivity, devices employing electrical readout such as electrically detected magnetic resonance (EDMR) do not require optics, making them easier to integrate with electronic circuitry and benefit from reduced size, weight, and power (SWaP). 

In EDMR, an applied bias generates a non-equilibrium population of electrons and holes in the active region of the device. In subsequent electron-hole recombination which produces a spin-dependent recombination (SDR) current, a carrier is first captured by a shallow level defect and then by a deep level defect. Depending on the resulting angular momentum selection rules a Pauli-spin blockade occurs, bottlenecking the SDR current, which is relieved when one of the paired spins is resonantly flipping. This is accomplished with the application of a quasi-static magnetic field, which brings the energy levels of one or both spins into resonance with an applied oscillating magnetic field, relieving the bottleneck and producing a characteristic change in the measured SDR current. A detailed and fully-quantitative theory of the recombination mechanism of silicon vacancy-related spin pairs is crucial to optimizing SiC power electronics and EDMR-based magnetometer technology.

\begin{figure}[t!]
\centering
\subfloat[]{
\includegraphics[width=0.49\linewidth]{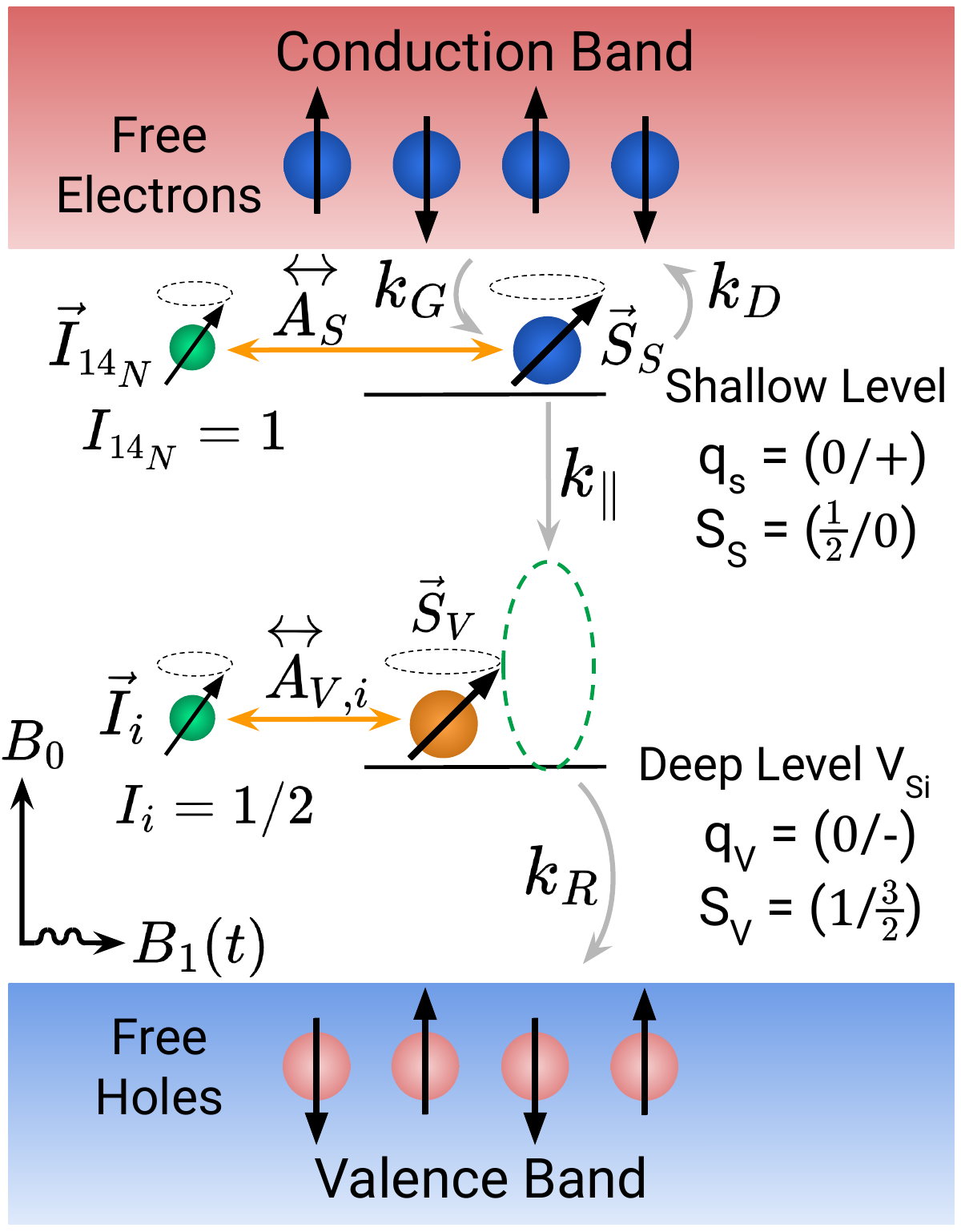}}
\subfloat[]{
\includegraphics[width=0.49\linewidth]{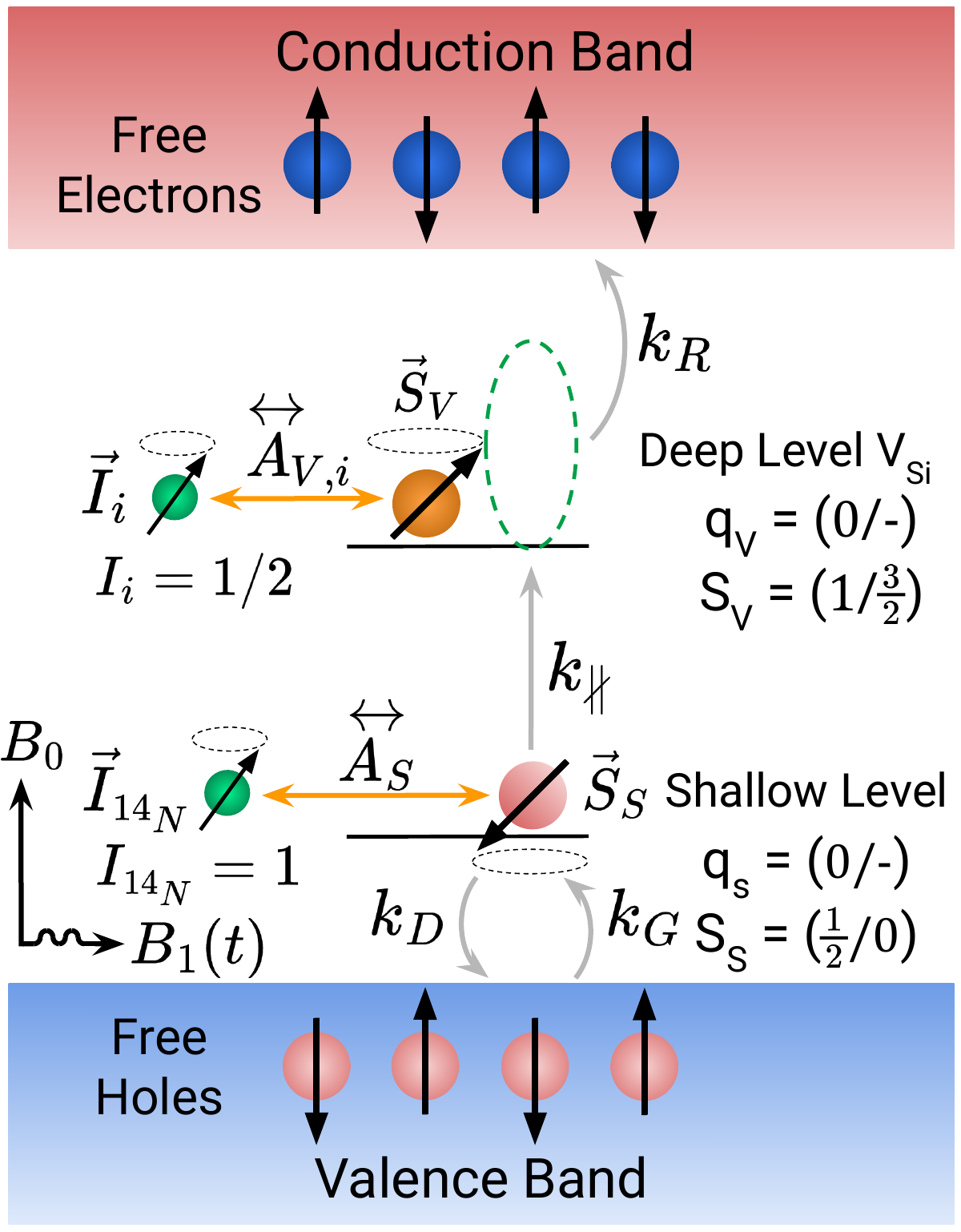}}
\caption{EDMR diagram of silicon vacancy-related spin pairs in 4H-SiC. The silicon vacancy is in the (0/$-$) charge and (1\big/3/2) spin state level, and interacts with a a dilute bath of proximal $^{29}$Si and $^{13}$C nuclear spins of nuclear spin magnetic moment $I_{i}=1/2$. The shallow level spin center is $S=1/2$ when neutrally charged and $S=0$ when ionized, and interacts with a single onsite $^{14}$N nuclear spin. (a) \underline{Shallow donor model}. Non-spin-polarized conduction electrons are captured by the shallow donor, generating the spin pair at a rate $k_{G}$. From there, the unpaired electron can tunnel to the neutrally-charged silicon vacancy at a rate $k_{\parallel}$ when the two spins are in a parallel configuration. Otherwise, the spin pair dissociates by depopulating the shallow level at a rate $k_{D}$. After the silicon vacancy becomes negatively-charged, it may release one electron to recombine with an anti-parallel-aligned valence hole at a rate $k_{R}$. \\(b) \underline{Shallow acceptor model}. Non-spin-polarized valence holes are captured by the shallow donor, generating the spin pair at a rate $k_{G}$. From there, the unpaired hole can tunnel to the negatively-charged silicon vacancy at a rate $k_{\nparallel}$ when the two spins are in an anti-parallel configuration. Otherwise, the spin pair dissociates by depopulating the shallow level at a rate $k_{D}$. After the silicon vacancy becomes neutrally-charged, it may release one hole to recombine with a parallel-aligned conduction electron at a rate $k_{R}$.}
\label{fig:EDMR_diagram}
\end{figure}

Here we present a quantitative theory for simulating EDMR measurements of silicon vacancy-related spin pairs in 4H-SiC and compare with experimental results from Ref. \onlinecite{CochraneCoreyJ.2016Vmfs}. We first show results for V1a-type silicon vacancies (zero-field splitting $\mathcal{D}\simeq0$ MHz), then V2a-type ($\mathcal{D}\simeq35$ MHz)\cite{PhysRevB.61.2613,PhysRevB.66.235202}, and end with a full mathematical description of the model. In our model, the silicon vacancy is near the $q_{V}=(0/-)$ charge state transition level\cite{CochraneC.J.2011Aedm}, corresponding to the spin quantum numbers $S_{V}^{[q]}=(1/\frac{3}{2})$. To make a charge transition from the neutral to negative state, it must accept an electron from a "nearby" (within $\sim10$~nm)\cite{Suckert01102013,PhysRevB.88.155301} shallow {\it donor} in the $q_{S}=(0/+)$ level, corresponding to the spin quantum numbers $S_{S}^{[q]}=(\frac{1}{2}/0)$. Thus, this transition is only allowed when the vacancy and donor electron spins are aligned \textbf{parallel}, as shown in Fig.~\ref{fig:EDMR_diagram}. On the other hand, the silicon vacancy could make a charge transition from the negative to neutral state by accepting a hole from a "nearby" shallow {\it acceptor} in the $q_{S}=(0/-)$ level, again corresponding to $S_{S}^{[q]}=(\frac{1}{2}/0)$. Thus, this transition is only allowed when the vacancy and acceptor hole spins are aligned \textbf{antiparallel}, as shown in Fig.~S1 in the Supplementary Information. The Hamiltonians of the shallow level spin center and the deep level silicon vacancy have the form:
\begin{align}\label{eq:Shallow Hamiltonian}
   \hat{H}_{S}&=g_{S}\mu_{B}\left(\vec{B}_{0}+\vec{B}_{1}(t)\right)\cdot\vec{S}_{S}+A_{S\parallel}\hat{S}_{S,z}\otimes\hat{I}_{S,z}
\end{align}
\begin{align}\label{eq:VSi Hamiltonian}
    \hat{H}_{V}^{[q]}&=\mathcal{D}^{[q]}\left(\left[\hat{S}_{V,z}^{[q]}\right]^{2}-\frac{1}{3}S_{V}^{[q]}(S_{V}^{[q]}+1)\hat{1}\right)+\hat{S}_{V,z}^{[q]}\otimes\sum A_{V\parallel,i}^{[q]}\hat{I}_{Vi,z}\notag\\&+g_{V}^{[q]}\mu_{B}\left(\vec{B}_{0}+\vec{B}_{1}(t)\right)\cdot\vec{S}_{V}^{[q]}
\end{align}
where $\hat{H}_{V}$ and $\hat{H}_{S}$ are the respective Hamiltonians of the silicon vacancy and the (occupied) shallow level spin center, $[q]$ denotes the charge state of the silicon vacancy, $\vec{S}$ is the vector of spin operators, $\mathcal{D}$ is the zero-field splitting, $\mu_{B}$ is the Bohr magneton, $\vec{B}_{0}$ is the applied magnetic field taken to be along $\hat{e}_{z}\parallel\hat{c}$, $|\vec{B}_{1}(t)|=B_{1}\cos(\omega t)$ is the microwave field taken to be along $\hat{e}_{x}$ with frequency $\omega$, $g_{V}$ and $g_{S}$ are the g-factors of the vacancy and shallow level spins, $A_{\parallel,k}$ are the hyperfine tensor components parallel to $\vec{B}_{0}$, and $\hat{I}_{k}$ are nuclear spin operators.

We simulate the EDMR spectrum of silicon vacancy-related spin pairs in 4H-SiC with a three-manifold model. For the shallow donor (acceptor) case, the first manifold tracks the dynamics of the neutral (negative) charged silicon vacancy while the shallow level is \textit{unoccupied}. The second manifold tracks the dynamics of the neutral (negative) charged silicon vacancy while the shallow level is \textit{occupied}. The third manifold tracks the dynamics of the negative (neutral) charged silicon vacancy while the shallow level is again unoccupied. We ignore any coherence that may evolve at the shallow level when the deep level is negative (neutral) charged. We define the Hamiltonians for each of these configurations in terms of $\hat{H}_{V}^{[q]}$ and $\hat{H}_{S}$ for the shallow donor case as $\hat{\mathcal{H}}_{1\otimes0}\equiv\hat{H}_{V}^{[0]}$, $\hat{\mathcal{H}}_{1\otimes1/2}\equiv\hat{H}_{V}^{[0]}\otimes\hat{1}_{1/2}+\hat{1}_{1}\otimes\hat{H}_{S}$, and $\hat{\mathcal{H}}_{3/2\otimes0}\equiv\hat{H}_{V}^{[-]}$, respectively, where the subscripts label the respective spins of the deep and shallow levels and the superscripts label the spinful charge states. Thus, the total Hamiltonian $\mathscr{H}(t)$ is block diagonal in $\hat{\mathcal{H}}_{1\otimes0}$, $\hat{\mathcal{H}}_{1\otimes1/2}$, and $\hat{\mathcal{H}}_{3/2\otimes0}$. Similarly, for the shallow acceptor case we have $\hat{\mathcal{H}}_{3/2\otimes0}\equiv\hat{H}_{V}^{[-]}$, $\hat{\mathcal{H}}_{3/2\otimes1/2}\equiv\hat{H}_{V}^{[-]}\otimes\hat{1}_{1/2}+\hat{1}_{3/2}\otimes\hat{H}_{S}$, and $\hat{\mathcal{H}}_{1\otimes0}\equiv\hat{H}_{V}^{[0]}$, and the total Hamiltonian $\mathscr{H}(t)$ is block diagonal in $\hat{\mathcal{H}}_{3/2\otimes0}$, $\hat{\mathcal{H}}_{3/2\otimes1/2}$, and $\hat{\mathcal{H}}_{1\otimes0}$. In either case, our simulations are most sensitive to the dynamics of $\hat{\mathcal{H}}_{1\otimes1/2}$ (donor model) and $\hat{\mathcal{H}}_{3/2\otimes1/2}$ (acceptor model).
To simulate the bath of $^{29}$Si and $^{13}$C nuclear spins ($I=1/2$ in 4.7$\%$ and 1.1$\%$ abundance, respectively) in the vicinity of the silicon vacancy, we calculate EDMR spectra of the most likely configurations and perform a weighted sum by their respective probabilities of occurrence. All combinatoric details can be found in the Supplementary Information.
\begin{figure}[t!]
\centering
\subfloat[]{
\includegraphics[width=0.53\linewidth]{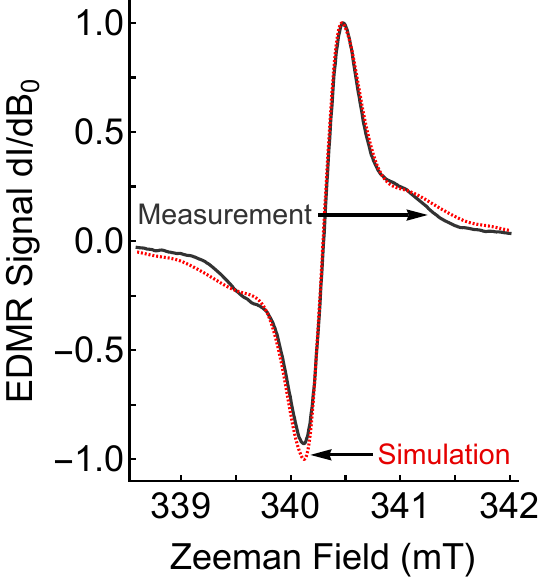}}\hspace{-0.3cm}
\subfloat[]{\includegraphics[width=0.46\linewidth]{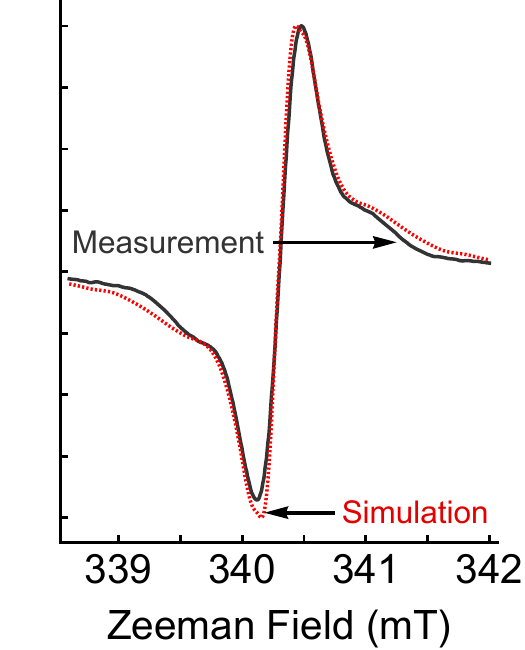}}
\caption{EDMR spectrum ($dI/dB_{0}$) of V1a silicon vacancy-related spin pairs in 4H-SiC normalized to the peak height versus magnetic field with a microwave frequency of 9.54 GHz. The simulated spectra (dashed red) is compared to the room-temperature measurement (solid black) in Ref.~\onlinecite{CochraneCoreyJ.2016Vmfs}. (a) Shallow donor model. (b) Shallow acceptor model, donor parameters $\times3/4$ and $k_{\parallel}\rightarrow k_{\nparallel}$. (Insets): Corresponding simulation and measurement\cite{CochraneCoreyJ.2016Vmfs} of the integrated EDMR spectrum $I(B_{0})$ versus magnetic field.}
\label{fig:RoomV1Fit}
\end{figure}

\begin{figure}[t!]
\centering
\subfloat[]{
\includegraphics[width=0.99\linewidth]{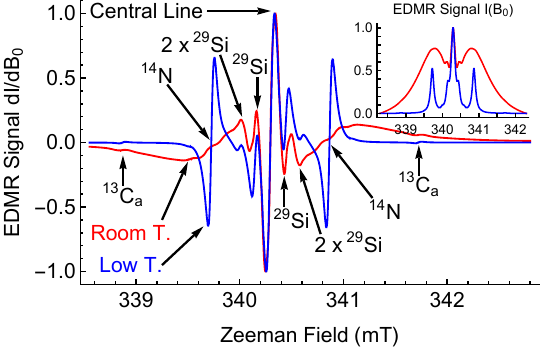}}\\
\subfloat[]{
\includegraphics[width=0.99\linewidth]{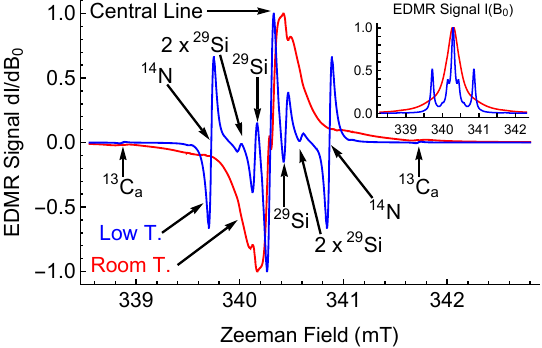}}
\caption{The predicted EDMR spectrum ($dI/dB_{0}$) of V1a silicon vacancy-related spin pairs in 4H-SiC normalized to the peak height at low microwave power, contrasting room temperature (red) and low temperature (blue) spectra. Features arising from hyperfine splitting are identified with arrows. (a) \underline{Shallow donor model}, $B_{1}=0.05$ mT. Low (Room) temperature parameters: $T_{1,S}=1\ \mu s$ (77 ns), $T_{2,S}=1\ \mu s$ (12.5 ns), $k_{\parallel}=1\ \mu s^{-1}$ (18 $\mu s^{-1}$). (b) \underline{Shallow acceptor model}, $B_{1}=0.0375$ mT. Low (Room) temperature parameters: $T_{1,S}=1\ \mu s$ (77 ns), $T_{2,S}=1\ \mu s$ (12.5 ns), $k_{\nparallel}=1\ \mu s^{-1}$ (13.5 $\mu s^{-1}$). (Insets): Corresponding simulation of the integrated EDMR spectrum $I(B_{0})$ versus magnetic field.}
\label{fig:V1smallB1}
\end{figure}

EDMR simulations from our theory of V1a silicon vacancy-related spin pairs in 4H-SiC are shown in Figures~\ref{fig:RoomV1Fit} and \ref{fig:V1smallB1} as functions of the applied Zeeman magnetic field and at a constant microwave frequency of $f=\omega/2\pi\simeq9.54$ GHz. Figure~\ref{fig:RoomV1Fit} shows the best-fit simulations to a room temperature EDMR measurement\cite{CochraneCoreyJ.2016Vmfs}, attributed to V1a silicon vacancy-related spin pairs in 4H-SiC. We use the same spin Hamiltonian parameters for both charge states of the V1a silicon vacancy as, to our knowledge, the literature lacks direct measurements of these quantities for the neutral charge state, namely\cite{CochraneC.J.2012Ioas,PhysRevB.66.235202,PhysRevB.68.165206}: $\mathcal{D}^{[0]}=\mathcal{D}^{[-]}=0$, $g_{V}^{[0]}=g_{V}^{[-]}=2.003$, $A_{\parallel,^{29}Si}^{[0]}=A_{\parallel,^{29}Si}^{[-]}=8.35$ MHz, $A_{\parallel,^{13}C_{b}}^{[0]}=A_{\parallel,^{13}C_{b}}^{[-]}=35$ MHz, and $A_{\parallel,^{13}C_{a}}^{[0]}=A_{\parallel,^{13}C_{a}}^{[-]}=80$ MHz. Importantly, we found that the hyperfine splitting of the silicon vacancy was not sufficient to produce the subtle shoulders in the measurement. Thus, we conclude that these features must arise from hyperfine splitting of the shallow level spin's spectrum, which must have a g-factor near 2.003 and a hyperfine coupling of $\simeq30.8$ MHz (if $I=1/2$) or $\simeq15.4$ MHz (if $I=1$) according to the analysis in Ref. \onlinecite{CochraneCoreyJ.2016Vmfs}. Importantly, these constraints on the shallow level spin Hamiltonian parameters preclude the simple substitutional centers \cite{https://doi.org/10.1002/1521,SonN.T.2010EaES,PhysRevB.64.085206,PhysRevB.70.193207,PhysRevB.64.085206,PhysRevB.70.193207,PhysRev.124.1083}: nitrogen, phosphorus, boron, and aluminum.

Previously, this spin pair has been linked to a $^{14}$N nuclear spin ($I=1$) in EDENDOR measurements\cite{WaskiewiczRyanJ.2019Eden}, and a recent EDMR measurement extracted the g-factor ($\simeq2.003$) and hyperfine coupling (0.57 mT $\simeq16$ MHz) between this nuclear spin and an unpaired electron spin \cite{HigaE.2020Edmr}. This EDMR spectrum was tentatively attributed to an ``incomplete K-center'' structure - a silicon dangling bond center with a single $^{14}$N substituting onto one of the basal carbon sites, whose charge nature (i.e. donor or acceptor) is unknown. In our theory, we ascribe the shoulders in the measurement to this same structure, as the hyperfine coupling and g-factor match the requirements exactly. $^{14}$N  can occur in 4H-SiC defect complexes as it is commonly used as an $n$-type dopant and can be introduced through nitric oxide anneals\cite{10.1063/1.4805355,10.1063/1.5045668,10.1063/1.3659689,10.1063/1.3131845}, the latter being the case for the sample measured in Ref.\onlinecite{CochraneCoreyJ.2016Vmfs}. Thus, for the shallow level spin Hamiltonian parameters we used $g_{S}=2.003$ and $A_{\parallel,^{14}N}=0.57$ mT $=16.0$ MHz. 

In our best fits in Figure~\ref{fig:RoomV1Fit} we left the coherence times of both spins, microwave field strength $B_{1}$, and the hopping rates as free parameters. For both models, we found $T_{1,S}=77$ ns, $T_{2,S}=12.5$ ns, and $T_{1,V}\geq T_{2,V}\geq 5\ \mu s$. For the shallow donor case, we found $B_{1}=0.14$ mT, $k_{\parallel}=18\ \mu s^{-1}$, and $k_{D}\leq10\ \mu s^{-1}$. For the shallow acceptor case, we found that the donor values multiplied by the ratio of the spin degeneracies, i.e., $(2S_{V}^{[0]}+1)/(2S_{V}^{[-]}+1)=3/4$, produced a good fit as this conserves the total parameter value through all spin channels. A table of all simulation parameters corresponding to each plot can be found in the Supplementary Information.

Figure~\ref{fig:V1smallB1} shows EDMR simulations of V1a silicon vacancy-related spin pairs at low microwave power, contrasted at room temperature and low temperature, exhibiting predicted resolved hyperfine splitting due to $^{13}$C, $^{29}$Si, and $^{14}$N nuclei. The "Central Line" is the overlapping spectra of the $P_{000}$ deep level configuration (no $^{13}$C or $^{29}$Si near the silicon vacancy) and the middle peak of the hyperfine-split shallow level spectrum. Donor model simulations for different values of $A_{\parallel,^{29}Si}^{[0]}$ and for varying Gaussian distributions of $k_{\parallel}$ are shown in Figures 1 and 2 in the Supplementary Information. Our theory predicts that, while resolving the hyperfine structure of the silicon vacancy may be possible at room temperature, the shallow level hyperfine structure may only be resolvable at lower temperatures due to the short coherence times. 
\begin{figure}[t!]
\centering
\subfloat[]{
\includegraphics[width=0.99\linewidth]{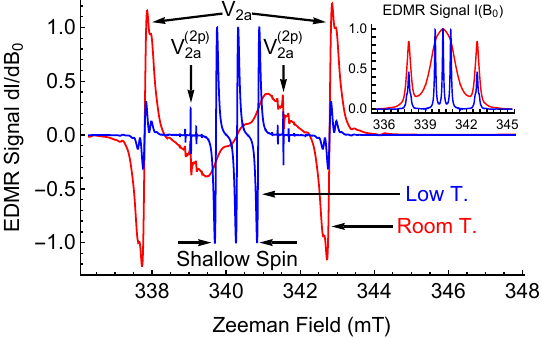}}\\
\subfloat[]{
\includegraphics[width=0.99\linewidth]{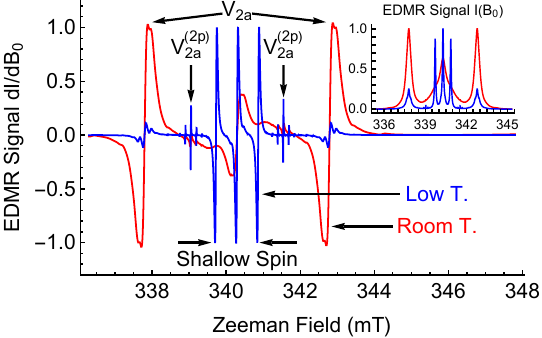}}
\caption{The predicted EDMR spectrum ($dI/dB_{0}$) of V2a silicon vacancy-related spin pairs in 4H-SiC normalized to the peak height at low microwave power, contrasting room temperature (red) and low temperature (blue) spectra. Features arising from each spin in the pair are identified with arrows. Similar simulation parameters were used here as in Figure~\ref{fig:V1smallB1}, except for $\mathcal{D}^{[q]}\neq0$. (a) Shallow donor model. (b) Shallow acceptor model. (Insets): Corresponding simulation of the integrated EDMR spectrum $I(B_{0})$ versus magnetic field.}
\label{fig:V2varParams}
\end{figure}

Figure~\ref{fig:V2varParams} illustrates the EDMR spectrum of V2a silicon vacancy-related spin pairs at low microwave power, contrasted at room temperature and low temperature, exhibiting the same hyperfine structure of both spin centers as in Figure~\ref{fig:V1smallB1}. We take similar simulation parameters as in Figure~\ref{fig:V1smallB1} but now with $\mathcal{D}^{[0]}=70$ MHz and $\mathcal{D}^{[-]}=35$ MHz, which gives the same zero field splitting energy for both charge states. Donor model simulations for different values of $\mathcal{D}^{[0]}$ are shown in Figure 3 in the Supplementary Information. Notably, the simulations in Figure~\ref{fig:V2varParams} predict that two-photon, double quantum transitions ($\mathrm{V_{2a}^{(2p)}}$) between $|\pm3/2\rangle\leftrightarrow|\mp1/2\rangle$ states of the negatively-charged silicon vacancy may be read-out electrically in either model. While optical read-out of these double-quantum transitions has been demonstrated \cite{KrausH.2014Rqme,PhysRevResearch.4.023022}, electrical read-out would provide an avenue to enhance magnetic field sensitivity \cite{PhysRevB.84.195204,RevModPhys.92.015004} in all-electric magnetometers \cite{CochraneCoreyJ.2016Vmfs,BakerW.J.2012Ramw}.

We now present the details of our theory for simulating EDMR measurements of silicon vacancy-related spin pairs. In Figs.~\ref{fig:RoomV1Fit}-\ref{fig:V2varParams} we modeled the EDMR dynamics with a Lindblad master equation\cite{GoriniVittorio1976Cpds,LindbladG.1976Otgo} of the form:
\begin{align}\label{eq:Master Equation}
       \partial_{t}\hat{\rho}(t)&=-\frac{i}{\hbar}\left[\hat{\mathscr{H}}(t),\hat{\rho}(t)\right]+\sum_{i}\hat{\mathcal{L}}_{hop.}\left[\hat{L}_{i}\right]+\sum_{i}\hat{\mathcal{L}}_{dec.}\left[\hat{L}_{i}\right]\notag\\ \hat{\mathcal{L}}\left[\hat{L}_{i}\right]&\equiv k_{i}\left(\hat{L}_{i}\hat{\rho}(t)\hat{L}_{i}^{\dagger}-\frac{1}{2}\left\{\hat{L}_{i}^{\dagger}\hat{L}_{i},\hat{\rho}(t)\right\}\right)
\end{align}
where $\hat{\mathscr{H}}(t)$ is block diagonal in $\hat{\mathcal{H}}_{1\otimes0}$, $\hat{\mathcal{H}}_{1\otimes1/2}$, and $\hat{\mathcal{H}}_{3/2\otimes0}$ (donor model) or $\hat{\mathcal{H}}_{3/2\otimes0}$, $\hat{\mathcal{H}}_{3/2\otimes1/2}$, and $\hat{\mathcal{H}}_{1\otimes0}$ (acceptor model); $\hat{\mathcal{L}}_{hop.}$ are the Lindblad superoperators for hopping processes in Figures~\ref{fig:EDMR_diagram} and S1; $\hat{\mathcal{L}}_{dec.}$ are the Lindblad superoperators for onsite decoherence; and $\hat{L}_{i}$ are the jump operators with respective rates $k_{i}$. Calculations are done in the rotating frame after performing the rotating wave approximation for computational ease. Additional calculation details are provided in the Supplementary Information.

The steady-state recombination current is determined by the density matrix populations of the manifold involved in recombination, i.e. the configuration corresponding to $\hat{\mathcal{H}}_{3/2\otimes0}$ (donor model) or $\hat{\mathcal{H}}_{1\otimes0}$ (acceptor model). This is achieved with the following current operator and steady-state current:
\begin{align}
    \begin{matrix}
        \hat{I}\equiv e\sum_{j}k_{R,j}\hat{L}^{\dagger}_{R,j}\hat{L}_{R,j}&&I(B_{0})\equiv Tr\left[\hat{I}\hat{\rho}_{ss}\right]
    \end{matrix}
    \label{eq:current operator}
\end{align}
where e is the elementary charge, $k_{R,j}$ and $\hat{L}_{R,j}$ are the recombination rates and jump operators, and $\hat{\rho}_{ss}$ is the steady-state solution of Eq.~\eqref{eq:Master Equation}. The current operator in Eq.~\eqref{eq:current operator} has the effect of tracing over the total population in the $\rho_{3/2\otimes0}$ manifold (donor model) or $\rho_{1\otimes0}$ (acceptor model).

This work presents a vital step toward optimizing SiC power electronics and EDMR-based magnetometer technology with a fully-quantitative theory of the recombination mechanism of silicon vacancy-related spin pairs in EDMR measurements using steady-state Lindblad equations. By fitting our model to experimental data attributed to V1a silicon vacancies, we constrained spin coherence times and electrical transport rates. Furthermore, we identified the shallow level spin center as a nitrogen-related complex and extracted its room-temperature coherence times. We also predicted how EDMR measurements of V1a and V2a silicon vacancy-related spin pairs are expected to respond to reduced temperatures. We conclude that while the silicon vacancy hyperfine structure may be resolvable at room temperature and at low microwave power, the shallow level hyperfine splitting may only emerge at lower temperatures due to its short coherence times. Finally, through EDMR simulations of V2a silicon vacancy-related spin pairs, we predict the emergence of two-photon double-quantum transitions which would offer an avenue towards all-electric enhanced magnetic field sensing. 

This work also advances previous theories\cite{10.1063/5.0172275,9039723} of EDMR by generalizing the spin-selective jump operators for $S>1/2$, widening the applicability of these theories to include spin pairs in most electrically-active charge states. Finally, this work shows how the effects of finite coherence times can be treated with the Lindblad formalism for $S=1/2$, 1, and $3/2$, which are vital to the simulation of other spin-dependent readout techniques such as optically detected magnetic resonance\cite{PatelRajN.2024RTDo,SadiMohammadAbdullah2025SEiS}, near-zero field magnetoresistance\cite{10.1063/5.0172275,10.1116/6.0003855,9039723}, and spin-polarized transport\cite{PhysRevLett.125.257203} measurements under realistic noise conditions. 

\begin{acknowledgments}
This material is based upon work supported by AFOSR FA9550-22-1-0308 and the NASA SC Space Grant Consortium. A portion of this research was carried out at the Jet Propulsion Laboratory, California Institute of Technology, under contract with the National Aeronautics and Space Administration, (contract 80NM0018D0004).
\end{acknowledgments}

\section*{Data Availability Statement}

The data that support the findings of this study are included in this published article and its supplementary information files.

\section*{References}

\providecommand{\noopsort}[1]{}\providecommand{\singleletter}[1]{#1}%


\begin{thebibliography}{44}%
\makeatletter
\providecommand \@ifxundefined [1]{%
 \@ifx{#1\undefined}
}%
\providecommand \@ifnum [1]{%
 \ifnum #1\expandafter \@firstoftwo
 \else \expandafter \@secondoftwo
 \fi
}%
\providecommand \@ifx [1]{%
 \ifx #1\expandafter \@firstoftwo
 \else \expandafter \@secondoftwo
 \fi
}%
\providecommand \natexlab [1]{#1}%
\providecommand \enquote  [1]{``#1''}%
\providecommand \bibnamefont  [1]{#1}%
\providecommand \bibfnamefont [1]{#1}%
\providecommand \citenamefont [1]{#1}%
\providecommand \href@noop [0]{\@secondoftwo}%
\providecommand \href [0]{\begingroup \@sanitize@url \@href}%
\providecommand \@href[1]{\@@startlink{#1}\@@href}%
\providecommand \@@href[1]{\endgroup#1\@@endlink}%
\providecommand \@sanitize@url [0]{\catcode `\\12\catcode `\$12\catcode
  `\&12\catcode `\#12\catcode `\^12\catcode `\_12\catcode `\%12\relax}%
\providecommand \@@startlink[1]{}%
\providecommand \@@endlink[0]{}%
\providecommand \url  [0]{\begingroup\@sanitize@url \@url }%
\providecommand \@url [1]{\endgroup\@href {#1}{\urlprefix }}%
\providecommand \urlprefix  [0]{URL }%
\providecommand \Eprint [0]{\href }%
\providecommand \doibase [0]{http://dx.doi.org/}%
\providecommand \selectlanguage [0]{\@gobble}%
\providecommand \bibinfo  [0]{\@secondoftwo}%
\providecommand \bibfield  [0]{\@secondoftwo}%
\providecommand \translation [1]{[#1]}%
\providecommand \BibitemOpen [0]{}%
\providecommand \bibitemStop [0]{}%
\providecommand \bibitemNoStop [0]{.\EOS\space}%
\providecommand \EOS [0]{\spacefactor3000\relax}%
\providecommand \BibitemShut  [1]{\csname bibitem#1\endcsname}%
\let\auto@bib@innerbib\@empty
\bibitem [{\citenamefont {Alves}\ \emph {et~al.}(2017)\citenamefont {Alves},
  \citenamefont {Gomes}, \citenamefont {Lefranc}, \citenamefont {De~A.~Pegado},
  \citenamefont {Jeannin}, \citenamefont {Luciano},\ and\ \citenamefont
  {Rocha}}]{8257396}%
  \BibitemOpen
  \bibfield  {author} {\bibinfo {author} {\bibfnamefont {L.~F.~S.}\
  \bibnamefont {Alves}}, \bibinfo {author} {\bibfnamefont {R.~C.~M.}\
  \bibnamefont {Gomes}}, \bibinfo {author} {\bibfnamefont {P.}~\bibnamefont
  {Lefranc}}, \bibinfo {author} {\bibfnamefont {R.}~\bibnamefont
  {De~A.~Pegado}}, \bibinfo {author} {\bibfnamefont {P.-O.}\ \bibnamefont
  {Jeannin}}, \bibinfo {author} {\bibfnamefont {B.}~\bibnamefont {Luciano}}, \
  and\ \bibinfo {author} {\bibfnamefont {F.~V.}\ \bibnamefont {Rocha}},\
  }\bibfield  {title} {\enquote {\bibinfo {title} {{SIC} power devices in power
  electronics: An overview},}\ }in\ \href {\doibase 10.1109/COBEP.2017.8257396}
  {\emph {\bibinfo {booktitle} {2017 Brazilian Power Electronics Conference
  (COBEP)}}}\ (\bibinfo {year} {2017})\ pp.\ \bibinfo {pages}
  {1--8}\BibitemShut {NoStop}%
\bibitem [{\citenamefont {Cochrane}, \citenamefont {Lenahan},\ and\
  \citenamefont {Lelis}(2007)}]{10.1063/1.2714285}%
  \BibitemOpen
  \bibfield  {author} {\bibinfo {author} {\bibfnamefont {C.~J.}\ \bibnamefont
  {Cochrane}}, \bibinfo {author} {\bibfnamefont {P.~M.}\ \bibnamefont
  {Lenahan}}, \ and\ \bibinfo {author} {\bibfnamefont {A.~J.}\ \bibnamefont
  {Lelis}},\ }\bibfield  {title} {\enquote {\bibinfo {title} {Deep level
  defects which limit current gain in 4{H} {S}i{C} bipolar junction
  transistors},}\ }\href {\doibase 10.1063/1.2714285} {\bibfield  {journal}
  {\bibinfo  {journal} {Applied Physics Letters}\ }\textbf {\bibinfo {volume}
  {90}},\ \bibinfo {pages} {123501} (\bibinfo {year} {2007})}\BibitemShut
  {NoStop}%
\bibitem [{\citenamefont {Cochrane}, \citenamefont {Lenahan},\ and\
  \citenamefont {Lelis}(2011{\natexlab{a}})}]{10.1063/1.3530600}%
  \BibitemOpen
  \bibfield  {author} {\bibinfo {author} {\bibfnamefont {C.~J.}\ \bibnamefont
  {Cochrane}}, \bibinfo {author} {\bibfnamefont {P.~M.}\ \bibnamefont
  {Lenahan}}, \ and\ \bibinfo {author} {\bibfnamefont {A.~J.}\ \bibnamefont
  {Lelis}},\ }\bibfield  {title} {\enquote {\bibinfo {title} {An electrically
  detected magnetic resonance study of performance limiting defects in {S}i{C}
  metal oxide semiconductor field effect transistors},}\ }\href {\doibase
  10.1063/1.3530600} {\bibfield  {journal} {\bibinfo  {journal} {Journal of
  Applied Physics}\ }\textbf {\bibinfo {volume} {109}},\ \bibinfo {pages}
  {014506} (\bibinfo {year} {2011}{\natexlab{a}})}\BibitemShut {NoStop}%
\bibitem [{\citenamefont {Pearton}\ \emph {et~al.}(2021)\citenamefont
  {Pearton}, \citenamefont {Aitkaliyeva}, \citenamefont {Xian}, \citenamefont
  {Ren}, \citenamefont {Khachatrian}, \citenamefont {Ildefonso}, \citenamefont
  {Islam}, \citenamefont {Jafar~Rasel}, \citenamefont {Haque}, \citenamefont
  {Polyakov},\ and\ \citenamefont {Kim}}]{PeartonS.J.2021RDiW}%
  \BibitemOpen
  \bibfield  {author} {\bibinfo {author} {\bibfnamefont {S.~J.}\ \bibnamefont
  {Pearton}}, \bibinfo {author} {\bibfnamefont {A.}~\bibnamefont
  {Aitkaliyeva}}, \bibinfo {author} {\bibfnamefont {M.}~\bibnamefont {Xian}},
  \bibinfo {author} {\bibfnamefont {F.}~\bibnamefont {Ren}}, \bibinfo {author}
  {\bibfnamefont {A.}~\bibnamefont {Khachatrian}}, \bibinfo {author}
  {\bibfnamefont {A.}~\bibnamefont {Ildefonso}}, \bibinfo {author}
  {\bibfnamefont {Z.}~\bibnamefont {Islam}}, \bibinfo {author} {\bibfnamefont
  {M.~A.}\ \bibnamefont {Jafar~Rasel}}, \bibinfo {author} {\bibfnamefont
  {A.}~\bibnamefont {Haque}}, \bibinfo {author} {\bibfnamefont {A.~Y.}\
  \bibnamefont {Polyakov}}, \ and\ \bibinfo {author} {\bibfnamefont
  {J.}~\bibnamefont {Kim}},\ }\bibfield  {title} {\enquote {\bibinfo {title}
  {Review—{R}adiation {D}amage in {W}ide and {U}ltra-{W}ide {B}andgap
  {S}emiconductors},}\ }\href@noop {} {\bibfield  {journal} {\bibinfo
  {journal} {ECS Journal of Solid State Science and Technology}\ }\textbf
  {\bibinfo {volume} {10}},\ \bibinfo {pages} {55008} (\bibinfo {year}
  {2021})}\BibitemShut {NoStop}%
\bibitem [{\citenamefont {Widmann}\ \emph {et~al.}(2015)\citenamefont
  {Widmann}, \citenamefont {Lee}, \citenamefont {Rendler}, \citenamefont {Son},
  \citenamefont {Fedder}, \citenamefont {Paik}, \citenamefont {Yang},
  \citenamefont {Zhao}, \citenamefont {Yang}, \citenamefont {Booker},
  \citenamefont {Denisenko}, \citenamefont {Jamali}, \citenamefont
  {Momenzadeh}, \citenamefont {Gerhardt}, \citenamefont {Ohshima},
  \citenamefont {Gali}, \citenamefont {Janzén},\ and\ \citenamefont
  {Wrachtrup}}]{WidmannMatthias2015Ccos}%
  \BibitemOpen
  \bibfield  {author} {\bibinfo {author} {\bibfnamefont {M.}~\bibnamefont
  {Widmann}}, \bibinfo {author} {\bibfnamefont {S.-Y.}\ \bibnamefont {Lee}},
  \bibinfo {author} {\bibfnamefont {T.}~\bibnamefont {Rendler}}, \bibinfo
  {author} {\bibfnamefont {N.~T.}\ \bibnamefont {Son}}, \bibinfo {author}
  {\bibfnamefont {H.}~\bibnamefont {Fedder}}, \bibinfo {author} {\bibfnamefont
  {S.}~\bibnamefont {Paik}}, \bibinfo {author} {\bibfnamefont {L.-P.}\
  \bibnamefont {Yang}}, \bibinfo {author} {\bibfnamefont {N.}~\bibnamefont
  {Zhao}}, \bibinfo {author} {\bibfnamefont {S.}~\bibnamefont {Yang}}, \bibinfo
  {author} {\bibfnamefont {I.}~\bibnamefont {Booker}}, \bibinfo {author}
  {\bibfnamefont {A.}~\bibnamefont {Denisenko}}, \bibinfo {author}
  {\bibfnamefont {M.}~\bibnamefont {Jamali}}, \bibinfo {author} {\bibfnamefont
  {S.~A.}\ \bibnamefont {Momenzadeh}}, \bibinfo {author} {\bibfnamefont
  {I.}~\bibnamefont {Gerhardt}}, \bibinfo {author} {\bibfnamefont
  {T.}~\bibnamefont {Ohshima}}, \bibinfo {author} {\bibfnamefont
  {A.}~\bibnamefont {Gali}}, \bibinfo {author} {\bibfnamefont {E.}~\bibnamefont
  {Janzén}}, \ and\ \bibinfo {author} {\bibfnamefont {J.}~\bibnamefont
  {Wrachtrup}},\ }\bibfield  {title} {\enquote {\bibinfo {title} {Coherent
  control of single spins in silicon carbide at room temperature},}\
  }\href@noop {} {\bibfield  {journal} {\bibinfo  {journal} {Nature Materials}\
  }\textbf {\bibinfo {volume} {14}},\ \bibinfo {pages} {164--168} (\bibinfo
  {year} {2015})}\BibitemShut {NoStop}%
\bibitem [{\citenamefont {Tahara}\ \emph {et~al.}(2025)\citenamefont {Tahara},
  \citenamefont {Tamura}, \citenamefont {Toyama}, \citenamefont {Nakane},
  \citenamefont {Kutsuki}, \citenamefont {Yamazaki},\ and\ \citenamefont
  {Ohshima}}]{TaharaKosuke2025Qswd}%
  \BibitemOpen
  \bibfield  {author} {\bibinfo {author} {\bibfnamefont {K.}~\bibnamefont
  {Tahara}}, \bibinfo {author} {\bibfnamefont {S.-i.}\ \bibnamefont {Tamura}},
  \bibinfo {author} {\bibfnamefont {H.}~\bibnamefont {Toyama}}, \bibinfo
  {author} {\bibfnamefont {J.~J.}\ \bibnamefont {Nakane}}, \bibinfo {author}
  {\bibfnamefont {K.}~\bibnamefont {Kutsuki}}, \bibinfo {author} {\bibfnamefont
  {Y.}~\bibnamefont {Yamazaki}}, \ and\ \bibinfo {author} {\bibfnamefont
  {T.}~\bibnamefont {Ohshima}},\ }\bibfield  {title} {\enquote {\bibinfo
  {title} {Quantum sensing with duplex qubits of silicon vacancy centers in
  {S}i{C} at room temperature},}\ }\href@noop {} {\bibfield  {journal}
  {\bibinfo  {journal} {npj Quantum Information}\ }\textbf {\bibinfo {volume}
  {11}},\ \bibinfo {pages} {58} (\bibinfo {year} {2025})}\BibitemShut {NoStop}%
\bibitem [{\citenamefont {Kraus}\ \emph
  {et~al.}(2014{\natexlab{a}})\citenamefont {Kraus}, \citenamefont {Soltamov},
  \citenamefont {Riedel}, \citenamefont {Väth}, \citenamefont {Fuchs},
  \citenamefont {Sperlich}, \citenamefont {Baranov}, \citenamefont {Dyakonov},\
  and\ \citenamefont {Astakhov}}]{KrausH.2014Rqme}%
  \BibitemOpen
  \bibfield  {author} {\bibinfo {author} {\bibfnamefont {H.}~\bibnamefont
  {Kraus}}, \bibinfo {author} {\bibfnamefont {V.}~\bibnamefont {Soltamov}},
  \bibinfo {author} {\bibfnamefont {D.}~\bibnamefont {Riedel}}, \bibinfo
  {author} {\bibfnamefont {S.}~\bibnamefont {Väth}}, \bibinfo {author}
  {\bibfnamefont {F.}~\bibnamefont {Fuchs}}, \bibinfo {author} {\bibfnamefont
  {A.}~\bibnamefont {Sperlich}}, \bibinfo {author} {\bibfnamefont
  {P.}~\bibnamefont {Baranov}}, \bibinfo {author} {\bibfnamefont
  {V.}~\bibnamefont {Dyakonov}}, \ and\ \bibinfo {author} {\bibfnamefont
  {G.}~\bibnamefont {Astakhov}},\ }\bibfield  {title} {\enquote {\bibinfo
  {title} {Room-temperature quantum microwave emitters based on spin defects in
  silicon carbide},}\ }\href@noop {} {\bibfield  {journal} {\bibinfo  {journal}
  {Nature Physics}\ }\textbf {\bibinfo {volume} {10}},\ \bibinfo {pages}
  {157--162} (\bibinfo {year} {2014}{\natexlab{a}})}\BibitemShut {NoStop}%
\bibitem [{\citenamefont {Cochrane}\ \emph {et~al.}(2016)\citenamefont
  {Cochrane}, \citenamefont {Blacksberg}, \citenamefont {Anders},\ and\
  \citenamefont {Lenahan}}]{CochraneCoreyJ.2016Vmfs}%
  \BibitemOpen
  \bibfield  {author} {\bibinfo {author} {\bibfnamefont {C.~J.}\ \bibnamefont
  {Cochrane}}, \bibinfo {author} {\bibfnamefont {J.}~\bibnamefont
  {Blacksberg}}, \bibinfo {author} {\bibfnamefont {M.~A.}\ \bibnamefont
  {Anders}}, \ and\ \bibinfo {author} {\bibfnamefont {P.~M.}\ \bibnamefont
  {Lenahan}},\ }\bibfield  {title} {\enquote {\bibinfo {title} {Vectorized
  magnetometer for space applications using electrical readout of atomic scale
  defects in silicon carbide},}\ }\href@noop {} {\bibfield  {journal} {\bibinfo
   {journal} {Scientific Reports}\ }\textbf {\bibinfo {volume} {6}},\ \bibinfo
  {pages} {37077} (\bibinfo {year} {2016})}\BibitemShut {NoStop}%
\bibitem [{\citenamefont {Gottscholl}\ \emph {et~al.}(2024)\citenamefont
  {Gottscholl}, \citenamefont {Kraus}, \citenamefont {Aichinger},\ and\
  \citenamefont {Cochrane}}]{GottschollAndreas2024Eter}%
  \BibitemOpen
  \bibfield  {author} {\bibinfo {author} {\bibfnamefont {A.}~\bibnamefont
  {Gottscholl}}, \bibinfo {author} {\bibfnamefont {H.}~\bibnamefont {Kraus}},
  \bibinfo {author} {\bibfnamefont {T.}~\bibnamefont {Aichinger}}, \ and\
  \bibinfo {author} {\bibfnamefont {C.~J.}\ \bibnamefont {Cochrane}},\
  }\bibfield  {title} {\enquote {\bibinfo {title} {Enhancing the electrical
  readout of the spin-dependent recombination current in {S}i{C} {JFET}s for
  {EDMR} based magnetometry using a tandem (de-)modulation technique},}\
  }\href@noop {} {\bibfield  {journal} {\bibinfo  {journal} {Scientific
  Reports}\ }\textbf {\bibinfo {volume} {14}},\ \bibinfo {pages} {14283}
  (\bibinfo {year} {2024})}\BibitemShut {NoStop}%
\bibitem [{\citenamefont {Lew}\ \emph {et~al.}(2023)\citenamefont {Lew},
  \citenamefont {Sewani}, \citenamefont {Iwamoto}, \citenamefont {Ohshima},
  \citenamefont {McCallum},\ and\ \citenamefont {Johnson}}]{10.1063/5.0154382}%
  \BibitemOpen
  \bibfield  {author} {\bibinfo {author} {\bibfnamefont {C.~T.-K.}\
  \bibnamefont {Lew}}, \bibinfo {author} {\bibfnamefont {V.~K.}\ \bibnamefont
  {Sewani}}, \bibinfo {author} {\bibfnamefont {N.}~\bibnamefont {Iwamoto}},
  \bibinfo {author} {\bibfnamefont {T.}~\bibnamefont {Ohshima}}, \bibinfo
  {author} {\bibfnamefont {J.~C.}\ \bibnamefont {McCallum}}, \ and\ \bibinfo
  {author} {\bibfnamefont {B.~C.}\ \bibnamefont {Johnson}},\ }\bibfield
  {title} {\enquote {\bibinfo {title} {Enhanced magnetometry with an
  electrically detected spin defect ensemble in silicon carbide},}\ }\href
  {\doibase 10.1063/5.0154382} {\bibfield  {journal} {\bibinfo  {journal}
  {Applied Physics Letters}\ }\textbf {\bibinfo {volume} {122}},\ \bibinfo
  {pages} {234001} (\bibinfo {year} {2023})}\BibitemShut {NoStop}%
\bibitem [{\citenamefont {Lew}\ \emph {et~al.}(2024)\citenamefont {Lew},
  \citenamefont {Sewani}, \citenamefont {Iwamoto}, \citenamefont {Ohshima},
  \citenamefont {McCallum},\ and\ \citenamefont
  {Johnson}}]{PhysRevLett.132.146902}%
  \BibitemOpen
  \bibfield  {author} {\bibinfo {author} {\bibfnamefont {C.~T.-K.}\
  \bibnamefont {Lew}}, \bibinfo {author} {\bibfnamefont {V.~K.}\ \bibnamefont
  {Sewani}}, \bibinfo {author} {\bibfnamefont {N.}~\bibnamefont {Iwamoto}},
  \bibinfo {author} {\bibfnamefont {T.}~\bibnamefont {Ohshima}}, \bibinfo
  {author} {\bibfnamefont {J.~C.}\ \bibnamefont {McCallum}}, \ and\ \bibinfo
  {author} {\bibfnamefont {B.~C.}\ \bibnamefont {Johnson}},\ }\bibfield
  {title} {\enquote {\bibinfo {title} {All-{E}lectrical {R}eadout of
  {C}oherently {C}ontrolled {S}pins in {S}ilicon {C}arbide},}\ }\href {\doibase
  10.1103/PhysRevLett.132.146902} {\bibfield  {journal} {\bibinfo  {journal}
  {Phys. Rev. Lett.}\ }\textbf {\bibinfo {volume} {132}},\ \bibinfo {pages}
  {146902} (\bibinfo {year} {2024})}\BibitemShut {NoStop}%
\bibitem [{\citenamefont {Kraus}\ \emph
  {et~al.}(2014{\natexlab{b}})\citenamefont {Kraus}, \citenamefont {Soltamov},
  \citenamefont {Fuchs}, \citenamefont {Simin}, \citenamefont {Sperlich},
  \citenamefont {Baranov}, \citenamefont {Astakhov},\ and\ \citenamefont
  {Dyakonov}}]{KrausH.2014Mfat}%
  \BibitemOpen
  \bibfield  {author} {\bibinfo {author} {\bibfnamefont {H.}~\bibnamefont
  {Kraus}}, \bibinfo {author} {\bibfnamefont {V.}~\bibnamefont {Soltamov}},
  \bibinfo {author} {\bibfnamefont {F.}~\bibnamefont {Fuchs}}, \bibinfo
  {author} {\bibfnamefont {D.}~\bibnamefont {Simin}}, \bibinfo {author}
  {\bibfnamefont {A.}~\bibnamefont {Sperlich}}, \bibinfo {author}
  {\bibfnamefont {P.}~\bibnamefont {Baranov}}, \bibinfo {author} {\bibfnamefont
  {G.}~\bibnamefont {Astakhov}}, \ and\ \bibinfo {author} {\bibfnamefont
  {V.}~\bibnamefont {Dyakonov}},\ }\bibfield  {title} {\enquote {\bibinfo
  {title} {Magnetic field and temperature sensing with atomic-scale spin
  defects in silicon carbide},}\ }\href@noop {} {\bibfield  {journal} {\bibinfo
   {journal} {Scientific Reports}\ }\textbf {\bibinfo {volume} {4}},\ \bibinfo
  {pages} {5303} (\bibinfo {year} {2014}{\natexlab{b}})}\BibitemShut {NoStop}%
\bibitem [{\citenamefont {Lekavicius}\ \emph {et~al.}(2023)\citenamefont
  {Lekavicius}, \citenamefont {Carter}, \citenamefont {Pennachio},
  \citenamefont {White}, \citenamefont {Hajzus}, \citenamefont {Purdy},
  \citenamefont {Gaskill}, \citenamefont {Yeats},\ and\ \citenamefont
  {Myers-Ward}}]{PhysRevApplied.19.044086}%
  \BibitemOpen
  \bibfield  {author} {\bibinfo {author} {\bibfnamefont {I.}~\bibnamefont
  {Lekavicius}}, \bibinfo {author} {\bibfnamefont {S.}~\bibnamefont {Carter}},
  \bibinfo {author} {\bibfnamefont {D.}~\bibnamefont {Pennachio}}, \bibinfo
  {author} {\bibfnamefont {S.}~\bibnamefont {White}}, \bibinfo {author}
  {\bibfnamefont {J.}~\bibnamefont {Hajzus}}, \bibinfo {author} {\bibfnamefont
  {A.}~\bibnamefont {Purdy}}, \bibinfo {author} {\bibfnamefont
  {D.}~\bibnamefont {Gaskill}}, \bibinfo {author} {\bibfnamefont
  {A.}~\bibnamefont {Yeats}}, \ and\ \bibinfo {author} {\bibfnamefont
  {R.}~\bibnamefont {Myers-Ward}},\ }\bibfield  {title} {\enquote {\bibinfo
  {title} {Magnetometry {B}ased on {S}ilicon-{V}acancy {C}enters in
  {I}sotopically {P}urified $4{H}$-$\mathrm{SiC}$},}\ }\href {\doibase
  10.1103/PhysRevApplied.19.044086} {\bibfield  {journal} {\bibinfo  {journal}
  {Phys. Rev. Appl.}\ }\textbf {\bibinfo {volume} {19}},\ \bibinfo {pages}
  {044086} (\bibinfo {year} {2023})}\BibitemShut {NoStop}%
\bibitem [{\citenamefont {Yamazaki}\ \emph {et~al.}(2023)\citenamefont
  {Yamazaki}, \citenamefont {Masuyama}, \citenamefont {Kojima},\ and\
  \citenamefont {Ohshima}}]{PhysRevApplied.20.L031001}%
  \BibitemOpen
  \bibfield  {author} {\bibinfo {author} {\bibfnamefont {Y.}~\bibnamefont
  {Yamazaki}}, \bibinfo {author} {\bibfnamefont {Y.}~\bibnamefont {Masuyama}},
  \bibinfo {author} {\bibfnamefont {K.}~\bibnamefont {Kojima}}, \ and\ \bibinfo
  {author} {\bibfnamefont {T.}~\bibnamefont {Ohshima}},\ }\bibfield  {title}
  {\enquote {\bibinfo {title} {Highly {S}ensitive {T}emperature {S}ensing
  {U}sing the {S}ilicon {V}acancy in {S}ilicon {C}arbide by {S}imultaneously
  {R}esonated {O}ptically {D}etected {M}agnetic {R}esonance},}\ }\href
  {\doibase 10.1103/PhysRevApplied.20.L031001} {\bibfield  {journal} {\bibinfo
  {journal} {Phys. Rev. Appl.}\ }\textbf {\bibinfo {volume} {20}},\ \bibinfo
  {pages} {L031001} (\bibinfo {year} {2023})}\BibitemShut {NoStop}%
\bibitem [{\citenamefont {S\"orman}\ \emph {et~al.}(2000)\citenamefont
  {S\"orman}, \citenamefont {Son}, \citenamefont {Chen}, \citenamefont
  {Kordina}, \citenamefont {Hallin},\ and\ \citenamefont
  {Janz\'en}}]{PhysRevB.61.2613}%
  \BibitemOpen
  \bibfield  {author} {\bibinfo {author} {\bibfnamefont {E.}~\bibnamefont
  {S\"orman}}, \bibinfo {author} {\bibfnamefont {N.~T.}\ \bibnamefont {Son}},
  \bibinfo {author} {\bibfnamefont {W.~M.}\ \bibnamefont {Chen}}, \bibinfo
  {author} {\bibfnamefont {O.}~\bibnamefont {Kordina}}, \bibinfo {author}
  {\bibfnamefont {C.}~\bibnamefont {Hallin}}, \ and\ \bibinfo {author}
  {\bibfnamefont {E.}~\bibnamefont {Janz\'en}},\ }\bibfield  {title} {\enquote
  {\bibinfo {title} {Silicon vacancy related defect in 4{H} and 6{H}
  {S}i{C}},}\ }\href {\doibase 10.1103/PhysRevB.61.2613} {\bibfield  {journal}
  {\bibinfo  {journal} {Phys. Rev. B}\ }\textbf {\bibinfo {volume} {61}},\
  \bibinfo {pages} {2613--2620} (\bibinfo {year} {2000})}\BibitemShut {NoStop}%
\bibitem [{\citenamefont {Mizuochi}\ \emph {et~al.}(2002)\citenamefont
  {Mizuochi}, \citenamefont {Yamasaki}, \citenamefont {Takizawa}, \citenamefont
  {Morishita}, \citenamefont {Ohshima}, \citenamefont {Itoh},\ and\
  \citenamefont {Isoya}}]{PhysRevB.66.235202}%
  \BibitemOpen
  \bibfield  {author} {\bibinfo {author} {\bibfnamefont {N.}~\bibnamefont
  {Mizuochi}}, \bibinfo {author} {\bibfnamefont {S.}~\bibnamefont {Yamasaki}},
  \bibinfo {author} {\bibfnamefont {H.}~\bibnamefont {Takizawa}}, \bibinfo
  {author} {\bibfnamefont {N.}~\bibnamefont {Morishita}}, \bibinfo {author}
  {\bibfnamefont {T.}~\bibnamefont {Ohshima}}, \bibinfo {author} {\bibfnamefont
  {H.}~\bibnamefont {Itoh}}, \ and\ \bibinfo {author} {\bibfnamefont
  {J.}~\bibnamefont {Isoya}},\ }\bibfield  {title} {\enquote {\bibinfo {title}
  {Continuous-wave and pulsed {EPR} study of the negatively charged silicon
  vacancy with ${S}=\frac{3}{2}$ and ${C}_{3v}$ symmetry in $n$-type
  $4{H}\ensuremath{-}\mathrm{SiC}$},}\ }\href {\doibase
  10.1103/PhysRevB.66.235202} {\bibfield  {journal} {\bibinfo  {journal} {Phys.
  Rev. B}\ }\textbf {\bibinfo {volume} {66}},\ \bibinfo {pages} {235202}
  (\bibinfo {year} {2002})}\BibitemShut {NoStop}%
\bibitem [{\citenamefont {Cochrane}, \citenamefont {Lenahan},\ and\
  \citenamefont {Lelis}(2011{\natexlab{b}})}]{CochraneC.J.2011Aedm}%
  \BibitemOpen
  \bibfield  {author} {\bibinfo {author} {\bibfnamefont {C.~J.}\ \bibnamefont
  {Cochrane}}, \bibinfo {author} {\bibfnamefont {P.~M.}\ \bibnamefont
  {Lenahan}}, \ and\ \bibinfo {author} {\bibfnamefont {A.~J.}\ \bibnamefont
  {Lelis}},\ }\bibfield  {title} {\enquote {\bibinfo {title} {An electrically
  detected magnetic resonance study of performance limiting defects in {S}i{C}
  metal oxide semiconductor field effect transistors},}\ }\href@noop {}
  {\bibfield  {journal} {\bibinfo  {journal} {Journal of Applied Physics}\
  }\textbf {\bibinfo {volume} {109}},\ \bibinfo {pages} {014506} (\bibinfo
  {year} {2011}{\natexlab{b}})}\BibitemShut {NoStop}%
\bibitem [{\citenamefont {Suckert}\ \emph {et~al.}(2013)\citenamefont
  {Suckert}, \citenamefont {Hoehne}, \citenamefont {Dreher}, \citenamefont
  {Kuenzl}, \citenamefont {Huebl}, \citenamefont {Stutzmann},\ and\
  \citenamefont {Brandt}}]{Suckert01102013}%
  \BibitemOpen
  \bibfield  {author} {\bibinfo {author} {\bibfnamefont {M.}~\bibnamefont
  {Suckert}}, \bibinfo {author} {\bibfnamefont {F.}~\bibnamefont {Hoehne}},
  \bibinfo {author} {\bibfnamefont {L.}~\bibnamefont {Dreher}}, \bibinfo
  {author} {\bibfnamefont {M.}~\bibnamefont {Kuenzl}}, \bibinfo {author}
  {\bibfnamefont {H.}~\bibnamefont {Huebl}}, \bibinfo {author} {\bibfnamefont
  {M.}~\bibnamefont {Stutzmann}}, \ and\ \bibinfo {author} {\bibfnamefont
  {M.~S.}\ \bibnamefont {Brandt}},\ }\bibfield  {title} {\enquote {\bibinfo
  {title} {Electrically detected double electron–electron resonance: exchange
  interaction of 31{P} donors and {P}b0 defects at the {S}i/{S}i{O}2
  interface},}\ }\href {\doibase 10.1080/00268976.2013.816796} {\bibfield
  {journal} {\bibinfo  {journal} {Molecular Physics}\ }\textbf {\bibinfo
  {volume} {111}},\ \bibinfo {pages} {2690--2695} (\bibinfo {year}
  {2013})}\BibitemShut {NoStop}%
\bibitem [{\citenamefont {Hoehne}\ \emph {et~al.}(2013)\citenamefont {Hoehne},
  \citenamefont {Dreher}, \citenamefont {Suckert}, \citenamefont {Franke},
  \citenamefont {Stutzmann},\ and\ \citenamefont
  {Brandt}}]{PhysRevB.88.155301}%
  \BibitemOpen
  \bibfield  {author} {\bibinfo {author} {\bibfnamefont {F.}~\bibnamefont
  {Hoehne}}, \bibinfo {author} {\bibfnamefont {L.}~\bibnamefont {Dreher}},
  \bibinfo {author} {\bibfnamefont {M.}~\bibnamefont {Suckert}}, \bibinfo
  {author} {\bibfnamefont {D.~P.}\ \bibnamefont {Franke}}, \bibinfo {author}
  {\bibfnamefont {M.}~\bibnamefont {Stutzmann}}, \ and\ \bibinfo {author}
  {\bibfnamefont {M.~S.}\ \bibnamefont {Brandt}},\ }\bibfield  {title}
  {\enquote {\bibinfo {title} {Time constants of spin-dependent recombination
  processes},}\ }\href {\doibase 10.1103/PhysRevB.88.155301} {\bibfield
  {journal} {\bibinfo  {journal} {Phys. Rev. B}\ }\textbf {\bibinfo {volume}
  {88}},\ \bibinfo {pages} {155301} (\bibinfo {year} {2013})}\BibitemShut
  {NoStop}%
\bibitem [{\citenamefont {Cochrane}, \citenamefont {Lenahan},\ and\
  \citenamefont {Lelis}(2012)}]{CochraneC.J.2012Ioas}%
  \BibitemOpen
  \bibfield  {author} {\bibinfo {author} {\bibfnamefont {C.~J.}\ \bibnamefont
  {Cochrane}}, \bibinfo {author} {\bibfnamefont {P.~M.}\ \bibnamefont
  {Lenahan}}, \ and\ \bibinfo {author} {\bibfnamefont {A.~J.}\ \bibnamefont
  {Lelis}},\ }\bibfield  {title} {\enquote {\bibinfo {title} {Identification of
  a silicon vacancy as an important defect in 4{H} {S}i{C} metal oxide
  semiconducting field effect transistor using spin dependent recombination},}\
  }\href@noop {} {\bibfield  {journal} {\bibinfo  {journal} {Applied Physics
  Letters}\ }\textbf {\bibinfo {volume} {100}},\ \bibinfo {pages} {023509}
  (\bibinfo {year} {2012})}\BibitemShut {NoStop}%
\bibitem [{\citenamefont {Mizuochi}\ \emph {et~al.}(2003)\citenamefont
  {Mizuochi}, \citenamefont {Yamasaki}, \citenamefont {Takizawa}, \citenamefont
  {Morishita}, \citenamefont {Ohshima}, \citenamefont {Itoh},\ and\
  \citenamefont {Isoya}}]{PhysRevB.68.165206}%
  \BibitemOpen
  \bibfield  {author} {\bibinfo {author} {\bibfnamefont {N.}~\bibnamefont
  {Mizuochi}}, \bibinfo {author} {\bibfnamefont {S.}~\bibnamefont {Yamasaki}},
  \bibinfo {author} {\bibfnamefont {H.}~\bibnamefont {Takizawa}}, \bibinfo
  {author} {\bibfnamefont {N.}~\bibnamefont {Morishita}}, \bibinfo {author}
  {\bibfnamefont {T.}~\bibnamefont {Ohshima}}, \bibinfo {author} {\bibfnamefont
  {H.}~\bibnamefont {Itoh}}, \ and\ \bibinfo {author} {\bibfnamefont
  {J.}~\bibnamefont {Isoya}},\ }\bibfield  {title} {\enquote {\bibinfo {title}
  {{EPR} studies of the isolated negatively charged silicon vacancies in n-type
  $4{H}$- and $6{H}$-{S}i{C}: {I}dentification of ${C}_{3v}$ symmetry and
  silicon sites},}\ }\href {\doibase 10.1103/PhysRevB.68.165206} {\bibfield
  {journal} {\bibinfo  {journal} {Phys. Rev. B}\ }\textbf {\bibinfo {volume}
  {68}},\ \bibinfo {pages} {165206} (\bibinfo {year} {2003})}\BibitemShut
  {NoStop}%
\bibitem [{\citenamefont
  {Greulich-Weber}(1997)}]{https://doi.org/10.1002/1521}%
  \BibitemOpen
  \bibfield  {author} {\bibinfo {author} {\bibfnamefont {S.}~\bibnamefont
  {Greulich-Weber}},\ }\bibfield  {title} {\enquote {\bibinfo {title} {{EPR}
  and {ENDOR} {I}nvestigations of {S}hallow {I}mpurities in {S}i{C}
  {P}olytypes},}\ }\href {\doibase
  https://doi.org/10.1002/1521-396X(199707)162:1<95::AID-PSSA95>3.0.CO;2-X}
  {\bibfield  {journal} {\bibinfo  {journal} {Physica Status Solidi (a)}\
  }\textbf {\bibinfo {volume} {162}},\ \bibinfo {pages} {95--151} (\bibinfo
  {year} {1997})}\BibitemShut {NoStop}%
\bibitem [{\citenamefont {Son}\ \emph {et~al.}(2010)\citenamefont {Son},
  \citenamefont {Isoya}, \citenamefont {Umeda}, \citenamefont {Ivanov},
  \citenamefont {Henry}, \citenamefont {Ohshima},\ and\ \citenamefont
  {Janzén}}]{SonN.T.2010EaES}%
  \BibitemOpen
  \bibfield  {author} {\bibinfo {author} {\bibfnamefont {N.~T.}\ \bibnamefont
  {Son}}, \bibinfo {author} {\bibfnamefont {J.}~\bibnamefont {Isoya}}, \bibinfo
  {author} {\bibfnamefont {T.}~\bibnamefont {Umeda}}, \bibinfo {author}
  {\bibfnamefont {I.~G.}\ \bibnamefont {Ivanov}}, \bibinfo {author}
  {\bibfnamefont {A.}~\bibnamefont {Henry}}, \bibinfo {author} {\bibfnamefont
  {T.}~\bibnamefont {Ohshima}}, \ and\ \bibinfo {author} {\bibfnamefont
  {E.}~\bibnamefont {Janzén}},\ }\bibfield  {title} {\enquote {\bibinfo
  {title} {{EPR} and {ENDOR} {S}tudies of {S}hallow {D}onors in {S}i{C}},}\
  }\href@noop {} {\bibfield  {journal} {\bibinfo  {journal} {Applied Magnetic
  Resonance}\ }\textbf {\bibinfo {volume} {39}},\ \bibinfo {pages} {49--85}
  (\bibinfo {year} {2010})}\BibitemShut {NoStop}%
\bibitem [{\citenamefont {v.~Duijn-Arnold}\ \emph {et~al.}(2001)\citenamefont
  {v.~Duijn-Arnold}, \citenamefont {Zondervan}, \citenamefont {Schmidt},
  \citenamefont {Baranov},\ and\ \citenamefont {Mokhov}}]{PhysRevB.64.085206}%
  \BibitemOpen
  \bibfield  {author} {\bibinfo {author} {\bibfnamefont {A.}~\bibnamefont
  {v.~Duijn-Arnold}}, \bibinfo {author} {\bibfnamefont {R.}~\bibnamefont
  {Zondervan}}, \bibinfo {author} {\bibfnamefont {J.}~\bibnamefont {Schmidt}},
  \bibinfo {author} {\bibfnamefont {P.~G.}\ \bibnamefont {Baranov}}, \ and\
  \bibinfo {author} {\bibfnamefont {E.~N.}\ \bibnamefont {Mokhov}},\ }\bibfield
   {title} {\enquote {\bibinfo {title} {Electronic structure of the {N} donor
  center in 4{H}-{S}i{C} and 6{H}-{S}i{C}},}\ }\href {\doibase
  10.1103/PhysRevB.64.085206} {\bibfield  {journal} {\bibinfo  {journal} {Phys.
  Rev. B}\ }\textbf {\bibinfo {volume} {64}},\ \bibinfo {pages} {085206}
  (\bibinfo {year} {2001})}\BibitemShut {NoStop}%
\bibitem [{\citenamefont {Son}\ \emph {et~al.}(2004)\citenamefont {Son},
  \citenamefont {Janz\'en}, \citenamefont {Isoya},\ and\ \citenamefont
  {Yamasaki}}]{PhysRevB.70.193207}%
  \BibitemOpen
  \bibfield  {author} {\bibinfo {author} {\bibfnamefont {N.~T.}\ \bibnamefont
  {Son}}, \bibinfo {author} {\bibfnamefont {E.}~\bibnamefont {Janz\'en}},
  \bibinfo {author} {\bibfnamefont {J.}~\bibnamefont {Isoya}}, \ and\ \bibinfo
  {author} {\bibfnamefont {S.}~\bibnamefont {Yamasaki}},\ }\bibfield  {title}
  {\enquote {\bibinfo {title} {Hyperfine interaction of the nitrogen donor in
  $4{H}\text{\ensuremath{-}}\mathrm{Si}\mathrm{C}$},}\ }\href {\doibase
  10.1103/PhysRevB.70.193207} {\bibfield  {journal} {\bibinfo  {journal} {Phys.
  Rev. B}\ }\textbf {\bibinfo {volume} {70}},\ \bibinfo {pages} {193207}
  (\bibinfo {year} {2004})}\BibitemShut {NoStop}%
\bibitem [{\citenamefont {Woodbury}\ and\ \citenamefont
  {Ludwig}(1961)}]{PhysRev.124.1083}%
  \BibitemOpen
  \bibfield  {author} {\bibinfo {author} {\bibfnamefont {H.~H.}\ \bibnamefont
  {Woodbury}}\ and\ \bibinfo {author} {\bibfnamefont {G.~W.}\ \bibnamefont
  {Ludwig}},\ }\bibfield  {title} {\enquote {\bibinfo {title} {Electron {S}pin
  {R}esonance {S}tudies in {S}i{C}},}\ }\href {\doibase
  10.1103/PhysRev.124.1083} {\bibfield  {journal} {\bibinfo  {journal} {Phys.
  Rev.}\ }\textbf {\bibinfo {volume} {124}},\ \bibinfo {pages} {1083--1089}
  (\bibinfo {year} {1961})}\BibitemShut {NoStop}%
\bibitem [{\citenamefont {Waskiewicz}\ \emph {et~al.}(2019)\citenamefont
  {Waskiewicz}, \citenamefont {Manning}, \citenamefont {McCrory},\ and\
  \citenamefont {Lenahan}}]{WaskiewiczRyanJ.2019Eden}%
  \BibitemOpen
  \bibfield  {author} {\bibinfo {author} {\bibfnamefont {R.~J.}\ \bibnamefont
  {Waskiewicz}}, \bibinfo {author} {\bibfnamefont {B.~R.}\ \bibnamefont
  {Manning}}, \bibinfo {author} {\bibfnamefont {D.~J.}\ \bibnamefont
  {McCrory}}, \ and\ \bibinfo {author} {\bibfnamefont {P.~M.}\ \bibnamefont
  {Lenahan}},\ }\bibfield  {title} {\enquote {\bibinfo {title} {Electrically
  detected electron nuclear double resonance in 4{H}-{S}i{C} bipolar junction
  transistors},}\ }\href {\doibase 10.1063/1.5108961} {\bibfield  {journal}
  {\bibinfo  {journal} {Journal of Applied Physics}\ }\textbf {\bibinfo
  {volume} {126}},\ \bibinfo {pages} {125709} (\bibinfo {year}
  {2019})}\BibitemShut {NoStop}%
\bibitem [{\citenamefont {Higa}\ \emph {et~al.}(2020)\citenamefont {Higa},
  \citenamefont {Sometani}, \citenamefont {Hirai}, \citenamefont {Yano},
  \citenamefont {Harada},\ and\ \citenamefont {Umeda}}]{HigaE.2020Edmr}%
  \BibitemOpen
  \bibfield  {author} {\bibinfo {author} {\bibfnamefont {E.}~\bibnamefont
  {Higa}}, \bibinfo {author} {\bibfnamefont {M.}~\bibnamefont {Sometani}},
  \bibinfo {author} {\bibfnamefont {H.}~\bibnamefont {Hirai}}, \bibinfo
  {author} {\bibfnamefont {H.}~\bibnamefont {Yano}}, \bibinfo {author}
  {\bibfnamefont {S.}~\bibnamefont {Harada}}, \ and\ \bibinfo {author}
  {\bibfnamefont {T.}~\bibnamefont {Umeda}},\ }\bibfield  {title} {\enquote
  {\bibinfo {title} {Electrically detected magnetic resonance study on
  interface defects at nitrided {S}i-face, a-face, and m-face
  4{H}-{S}i{C}/{S}i{O}2 interfaces},}\ }\href {\doibase 10.1063/5.0002944}
  {\bibfield  {journal} {\bibinfo  {journal} {Applied Physics Letters}\
  }\textbf {\bibinfo {volume} {116}},\ \bibinfo {pages} {171602} (\bibinfo
  {year} {2020})}\BibitemShut {NoStop}%
\bibitem [{\citenamefont {Cochrane}, \citenamefont {Lenahan},\ and\
  \citenamefont {Lelis}(2013)}]{10.1063/1.4805355}%
  \BibitemOpen
  \bibfield  {author} {\bibinfo {author} {\bibfnamefont {C.~J.}\ \bibnamefont
  {Cochrane}}, \bibinfo {author} {\bibfnamefont {P.~M.}\ \bibnamefont
  {Lenahan}}, \ and\ \bibinfo {author} {\bibfnamefont {A.~J.}\ \bibnamefont
  {Lelis}},\ }\bibfield  {title} {\enquote {\bibinfo {title} {The effect of
  nitric oxide anneals on silicon vacancies at and very near the interface of
  4{H} {S}i{C} metal oxide semiconducting field effect transistors using
  electrically detected magnetic resonance},}\ }\href {\doibase
  10.1063/1.4805355} {\bibfield  {journal} {\bibinfo  {journal} {Applied
  Physics Letters}\ }\textbf {\bibinfo {volume} {102}},\ \bibinfo {pages}
  {193507} (\bibinfo {year} {2013})}\BibitemShut {NoStop}%
\bibitem [{\citenamefont {Anders}\ \emph {et~al.}(2018)\citenamefont {Anders},
  \citenamefont {Lenahan}, \citenamefont {Edwards}, \citenamefont {Schultz},\
  and\ \citenamefont {Van~Ginhoven}}]{10.1063/1.5045668}%
  \BibitemOpen
  \bibfield  {author} {\bibinfo {author} {\bibfnamefont {M.~A.}\ \bibnamefont
  {Anders}}, \bibinfo {author} {\bibfnamefont {P.~M.}\ \bibnamefont {Lenahan}},
  \bibinfo {author} {\bibfnamefont {A.~H.}\ \bibnamefont {Edwards}}, \bibinfo
  {author} {\bibfnamefont {P.~A.}\ \bibnamefont {Schultz}}, \ and\ \bibinfo
  {author} {\bibfnamefont {R.~M.}\ \bibnamefont {Van~Ginhoven}},\ }\bibfield
  {title} {\enquote {\bibinfo {title} {Effects of nitrogen on the interface
  density of states distribution in 4{H}-{S}i{C} metal oxide semiconductor
  field effect transistors: {S}uper-hyperfine interactions and near interface
  silicon vacancy energy levels},}\ }\href {\doibase 10.1063/1.5045668}
  {\bibfield  {journal} {\bibinfo  {journal} {Journal of Applied Physics}\
  }\textbf {\bibinfo {volume} {124}},\ \bibinfo {pages} {184501} (\bibinfo
  {year} {2018})}\BibitemShut {NoStop}%
\bibitem [{\citenamefont {Kosugi}, \citenamefont {Umeda},\ and\ \citenamefont
  {Sakuma}(2011)}]{10.1063/1.3659689}%
  \BibitemOpen
  \bibfield  {author} {\bibinfo {author} {\bibfnamefont {R.}~\bibnamefont
  {Kosugi}}, \bibinfo {author} {\bibfnamefont {T.}~\bibnamefont {Umeda}}, \
  and\ \bibinfo {author} {\bibfnamefont {Y.}~\bibnamefont {Sakuma}},\
  }\bibfield  {title} {\enquote {\bibinfo {title} {Fixed nitrogen atoms in the
  {S}i{O}2/{S}i{C} interface region and their direct relationship to interface
  trap density},}\ }\href {\doibase 10.1063/1.3659689} {\bibfield  {journal}
  {\bibinfo  {journal} {Applied Physics Letters}\ }\textbf {\bibinfo {volume}
  {99}},\ \bibinfo {pages} {182111} (\bibinfo {year} {2011})}\BibitemShut
  {NoStop}%
\bibitem [{\citenamefont {Rozen}\ \emph {et~al.}(2009)\citenamefont {Rozen},
  \citenamefont {Dhar}, \citenamefont {Zvanut}, \citenamefont {Williams},\ and\
  \citenamefont {Feldman}}]{10.1063/1.3131845}%
  \BibitemOpen
  \bibfield  {author} {\bibinfo {author} {\bibfnamefont {J.}~\bibnamefont
  {Rozen}}, \bibinfo {author} {\bibfnamefont {S.}~\bibnamefont {Dhar}},
  \bibinfo {author} {\bibfnamefont {M.~E.}\ \bibnamefont {Zvanut}}, \bibinfo
  {author} {\bibfnamefont {J.~R.}\ \bibnamefont {Williams}}, \ and\ \bibinfo
  {author} {\bibfnamefont {L.~C.}\ \bibnamefont {Feldman}},\ }\bibfield
  {title} {\enquote {\bibinfo {title} {Density of interface states, electron
  traps, and hole traps as a function of the nitrogen density in {S}i{O}2 on
  {S}i{C}},}\ }\href {\doibase 10.1063/1.3131845} {\bibfield  {journal}
  {\bibinfo  {journal} {Journal of Applied Physics}\ }\textbf {\bibinfo
  {volume} {105}},\ \bibinfo {pages} {124506} (\bibinfo {year}
  {2009})}\BibitemShut {NoStop}%
\bibitem [{\citenamefont {Singh}\ \emph {et~al.}(2022)\citenamefont {Singh},
  \citenamefont {Hollberg}, \citenamefont {Anisimov}, \citenamefont {Baranov},\
  and\ \citenamefont {Suter}}]{PhysRevResearch.4.023022}%
  \BibitemOpen
  \bibfield  {author} {\bibinfo {author} {\bibfnamefont {H.}~\bibnamefont
  {Singh}}, \bibinfo {author} {\bibfnamefont {M.~A.}\ \bibnamefont {Hollberg}},
  \bibinfo {author} {\bibfnamefont {A.~N.}\ \bibnamefont {Anisimov}}, \bibinfo
  {author} {\bibfnamefont {P.~G.}\ \bibnamefont {Baranov}}, \ and\ \bibinfo
  {author} {\bibfnamefont {D.}~\bibnamefont {Suter}},\ }\bibfield  {title}
  {\enquote {\bibinfo {title} {Multi-photon multi-quantum transitions in the
  spin-$\frac{3}{2}$ silicon-vacancy centers of sic},}\ }\href {\doibase
  10.1103/PhysRevResearch.4.023022} {\bibfield  {journal} {\bibinfo  {journal}
  {Phys. Rev. Res.}\ }\textbf {\bibinfo {volume} {4}},\ \bibinfo {pages}
  {023022} (\bibinfo {year} {2022})}\BibitemShut {NoStop}%
\bibitem [{\citenamefont {Dr\'eau}\ \emph {et~al.}(2011)\citenamefont
  {Dr\'eau}, \citenamefont {Lesik}, \citenamefont {Rondin}, \citenamefont
  {Spinicelli}, \citenamefont {Arcizet}, \citenamefont {Roch},\ and\
  \citenamefont {Jacques}}]{PhysRevB.84.195204}%
  \BibitemOpen
  \bibfield  {author} {\bibinfo {author} {\bibfnamefont {A.}~\bibnamefont
  {Dr\'eau}}, \bibinfo {author} {\bibfnamefont {M.}~\bibnamefont {Lesik}},
  \bibinfo {author} {\bibfnamefont {L.}~\bibnamefont {Rondin}}, \bibinfo
  {author} {\bibfnamefont {P.}~\bibnamefont {Spinicelli}}, \bibinfo {author}
  {\bibfnamefont {O.}~\bibnamefont {Arcizet}}, \bibinfo {author} {\bibfnamefont
  {J.-F.}\ \bibnamefont {Roch}}, \ and\ \bibinfo {author} {\bibfnamefont
  {V.}~\bibnamefont {Jacques}},\ }\bibfield  {title} {\enquote {\bibinfo
  {title} {Avoiding power broadening in optically detected magnetic resonance
  of single {NV} defects for enhanced dc magnetic field sensitivity},}\ }\href
  {\doibase 10.1103/PhysRevB.84.195204} {\bibfield  {journal} {\bibinfo
  {journal} {Phys. Rev. B}\ }\textbf {\bibinfo {volume} {84}},\ \bibinfo
  {pages} {195204} (\bibinfo {year} {2011})}\BibitemShut {NoStop}%
\bibitem [{\citenamefont {Barry}\ \emph {et~al.}(2020)\citenamefont {Barry},
  \citenamefont {Schloss}, \citenamefont {Bauch}, \citenamefont {Turner},
  \citenamefont {Hart}, \citenamefont {Pham},\ and\ \citenamefont
  {Walsworth}}]{RevModPhys.92.015004}%
  \BibitemOpen
  \bibfield  {author} {\bibinfo {author} {\bibfnamefont {J.~F.}\ \bibnamefont
  {Barry}}, \bibinfo {author} {\bibfnamefont {J.~M.}\ \bibnamefont {Schloss}},
  \bibinfo {author} {\bibfnamefont {E.}~\bibnamefont {Bauch}}, \bibinfo
  {author} {\bibfnamefont {M.~J.}\ \bibnamefont {Turner}}, \bibinfo {author}
  {\bibfnamefont {C.~A.}\ \bibnamefont {Hart}}, \bibinfo {author}
  {\bibfnamefont {L.~M.}\ \bibnamefont {Pham}}, \ and\ \bibinfo {author}
  {\bibfnamefont {R.~L.}\ \bibnamefont {Walsworth}},\ }\bibfield  {title}
  {\enquote {\bibinfo {title} {Sensitivity optimization for {NV}-diamond
  magnetometry},}\ }\href {\doibase 10.1103/RevModPhys.92.015004} {\bibfield
  {journal} {\bibinfo  {journal} {Rev. Mod. Phys.}\ }\textbf {\bibinfo {volume}
  {92}},\ \bibinfo {pages} {015004} (\bibinfo {year} {2020})}\BibitemShut
  {NoStop}%
\bibitem [{\citenamefont {Baker}\ \emph {et~al.}(2012)\citenamefont {Baker},
  \citenamefont {Ambal}, \citenamefont {Waters}, \citenamefont {Baarda},
  \citenamefont {Morishita}, \citenamefont {van Schooten}, \citenamefont
  {McCamey}, \citenamefont {Lupton},\ and\ \citenamefont
  {Boehme}}]{BakerW.J.2012Ramw}%
  \BibitemOpen
  \bibfield  {author} {\bibinfo {author} {\bibfnamefont {W.}~\bibnamefont
  {Baker}}, \bibinfo {author} {\bibfnamefont {K.}~\bibnamefont {Ambal}},
  \bibinfo {author} {\bibfnamefont {D.}~\bibnamefont {Waters}}, \bibinfo
  {author} {\bibfnamefont {R.}~\bibnamefont {Baarda}}, \bibinfo {author}
  {\bibfnamefont {H.}~\bibnamefont {Morishita}}, \bibinfo {author}
  {\bibfnamefont {K.}~\bibnamefont {van Schooten}}, \bibinfo {author}
  {\bibfnamefont {D.}~\bibnamefont {McCamey}}, \bibinfo {author} {\bibfnamefont
  {J.}~\bibnamefont {Lupton}}, \ and\ \bibinfo {author} {\bibfnamefont
  {C.}~\bibnamefont {Boehme}},\ }\bibfield  {title} {\enquote {\bibinfo {title}
  {Robust absolute magnetometry with organic thin-film devices},}\ }\href@noop
  {} {\bibfield  {journal} {\bibinfo  {journal} {Nature Communications}\
  }\textbf {\bibinfo {volume} {3}},\ \bibinfo {pages} {898} (\bibinfo {year}
  {2012})}\BibitemShut {NoStop}%
\bibitem [{\citenamefont {Gorini}, \citenamefont {Kossakowski},\ and\
  \citenamefont {Sudarshan}(1976)}]{GoriniVittorio1976Cpds}%
  \BibitemOpen
  \bibfield  {author} {\bibinfo {author} {\bibfnamefont {V.}~\bibnamefont
  {Gorini}}, \bibinfo {author} {\bibfnamefont {A.}~\bibnamefont {Kossakowski}},
  \ and\ \bibinfo {author} {\bibfnamefont {E.~C.~G.}\ \bibnamefont
  {Sudarshan}},\ }\bibfield  {title} {\enquote {\bibinfo {title} {Completely
  positive dynamical semigroups of {N}‐level systems},}\ }\href@noop {}
  {\bibfield  {journal} {\bibinfo  {journal} {Journal of Mathematical Physics}\
  }\textbf {\bibinfo {volume} {17}},\ \bibinfo {pages} {821--825} (\bibinfo
  {year} {1976})}\BibitemShut {NoStop}%
\bibitem [{\citenamefont {Lindblad}(1976)}]{LindbladG.1976Otgo}%
  \BibitemOpen
  \bibfield  {author} {\bibinfo {author} {\bibfnamefont {G.}~\bibnamefont
  {Lindblad}},\ }\bibfield  {title} {\enquote {\bibinfo {title} {On the
  generators of quantum dynamical semigroups},}\ }\href@noop {} {\bibfield
  {journal} {\bibinfo  {journal} {Communications in Mathematical Physics}\
  }\textbf {\bibinfo {volume} {48}},\ \bibinfo {pages} {119--130} (\bibinfo
  {year} {1976})}\BibitemShut {NoStop}%
\bibitem [{\citenamefont {Harmon}\ \emph {et~al.}(2023)\citenamefont {Harmon},
  \citenamefont {Ashton}, \citenamefont {Lenahan},\ and\ \citenamefont
  {Flatté}}]{10.1063/5.0172275}%
  \BibitemOpen
  \bibfield  {author} {\bibinfo {author} {\bibfnamefont {N.~J.}\ \bibnamefont
  {Harmon}}, \bibinfo {author} {\bibfnamefont {J.~P.}\ \bibnamefont {Ashton}},
  \bibinfo {author} {\bibfnamefont {P.~M.}\ \bibnamefont {Lenahan}}, \ and\
  \bibinfo {author} {\bibfnamefont {M.~E.}\ \bibnamefont {Flatté}},\
  }\bibfield  {title} {\enquote {\bibinfo {title} {Spin-dependent capture
  mechanism for magnetic field effects on interface recombination current in
  semiconductor devices},}\ }\href {\doibase 10.1063/5.0172275} {\bibfield
  {journal} {\bibinfo  {journal} {Applied Physics Letters}\ }\textbf {\bibinfo
  {volume} {123}},\ \bibinfo {pages} {251603} (\bibinfo {year}
  {2023})}\BibitemShut {NoStop}%
\bibitem [{\citenamefont {Harmon}\ \emph {et~al.}(2020)\citenamefont {Harmon},
  \citenamefont {Mcmillan}, \citenamefont {Ashton}, \citenamefont {Lenahan},\
  and\ \citenamefont {Flatté}}]{9039723}%
  \BibitemOpen
  \bibfield  {author} {\bibinfo {author} {\bibfnamefont {N.~J.}\ \bibnamefont
  {Harmon}}, \bibinfo {author} {\bibfnamefont {S.~R.}\ \bibnamefont
  {Mcmillan}}, \bibinfo {author} {\bibfnamefont {J.~P.}\ \bibnamefont
  {Ashton}}, \bibinfo {author} {\bibfnamefont {P.~M.}\ \bibnamefont {Lenahan}},
  \ and\ \bibinfo {author} {\bibfnamefont {M.~E.}\ \bibnamefont {Flatté}},\
  }\bibfield  {title} {\enquote {\bibinfo {title} {Modeling of {N}ear
  {Z}ero-{F}ield {M}agnetoresistance and {E}lectrically {D}etected {M}agnetic
  {R}esonance in {I}rradiated {S}i/{S}i{O}2 {MOSFETs}},}\ }\href {\doibase
  10.1109/TNS.2020.2981495} {\bibfield  {journal} {\bibinfo  {journal} {IEEE
  Transactions on Nuclear Science}\ }\textbf {\bibinfo {volume} {67}},\
  \bibinfo {pages} {1669--1673} (\bibinfo {year} {2020})}\BibitemShut {NoStop}%
\bibitem [{\citenamefont {Patel}\ \emph {et~al.}(2024)\citenamefont {Patel},
  \citenamefont {Fishman}, \citenamefont {Huang}, \citenamefont {Gusdorff},
  \citenamefont {Fehr}, \citenamefont {Hopper}, \citenamefont {Breitweiser},
  \citenamefont {Porat}, \citenamefont {Flatt\'e},\ and\ \citenamefont
  {Bassett}}]{PatelRajN.2024RTDo}%
  \BibitemOpen
  \bibfield  {author} {\bibinfo {author} {\bibfnamefont {R.~N.}\ \bibnamefont
  {Patel}}, \bibinfo {author} {\bibfnamefont {R.~E.~K.}\ \bibnamefont
  {Fishman}}, \bibinfo {author} {\bibfnamefont {T.-Y.}\ \bibnamefont {Huang}},
  \bibinfo {author} {\bibfnamefont {J.~A.}\ \bibnamefont {Gusdorff}}, \bibinfo
  {author} {\bibfnamefont {D.~A.}\ \bibnamefont {Fehr}}, \bibinfo {author}
  {\bibfnamefont {D.~A.}\ \bibnamefont {Hopper}}, \bibinfo {author}
  {\bibfnamefont {S.~A.}\ \bibnamefont {Breitweiser}}, \bibinfo {author}
  {\bibfnamefont {B.}~\bibnamefont {Porat}}, \bibinfo {author} {\bibfnamefont
  {M.~E.}\ \bibnamefont {Flatt\'e}}, \ and\ \bibinfo {author} {\bibfnamefont
  {L.~C.}\ \bibnamefont {Bassett}},\ }\bibfield  {title} {\enquote {\bibinfo
  {title} {Room {T}emperature {D}ynamics of an {O}ptically {A}ddressable
  {S}ingle {S}pin in {H}exagonal {B}oron {N}itride},}\ }\href@noop {}
  {\bibfield  {journal} {\bibinfo  {journal} {Nano Letters}\ }\textbf {\bibinfo
  {volume} {24}},\ \bibinfo {pages} {7623--7628} (\bibinfo {year}
  {2024})}\BibitemShut {NoStop}%
\bibitem [{\citenamefont {Sadi}\ \emph {et~al.}(2025)\citenamefont {Sadi},
  \citenamefont {Basso}, \citenamefont {Fehr}, \citenamefont {Gao},
  \citenamefont {Vaidya}, \citenamefont {Riendeau}, \citenamefont {Joshi},
  \citenamefont {Li}, \citenamefont {Flatt\'e}, \citenamefont {Mounce},\ and\
  \citenamefont {Chen}}]{SadiMohammadAbdullah2025SEiS}%
  \BibitemOpen
  \bibfield  {author} {\bibinfo {author} {\bibfnamefont {M.~A.}\ \bibnamefont
  {Sadi}}, \bibinfo {author} {\bibfnamefont {L.}~\bibnamefont {Basso}},
  \bibinfo {author} {\bibfnamefont {D.~A.}\ \bibnamefont {Fehr}}, \bibinfo
  {author} {\bibfnamefont {X.}~\bibnamefont {Gao}}, \bibinfo {author}
  {\bibfnamefont {S.}~\bibnamefont {Vaidya}}, \bibinfo {author} {\bibfnamefont
  {E.~G.}\ \bibnamefont {Riendeau}}, \bibinfo {author} {\bibfnamefont
  {G.}~\bibnamefont {Joshi}}, \bibinfo {author} {\bibfnamefont
  {T.}~\bibnamefont {Li}}, \bibinfo {author} {\bibfnamefont {M.~E.}\
  \bibnamefont {Flatt\'e}}, \bibinfo {author} {\bibfnamefont {A.~M.}\
  \bibnamefont {Mounce}}, \ and\ \bibinfo {author} {\bibfnamefont {Y.~P.}\
  \bibnamefont {Chen}},\ }\bibfield  {title} {\enquote {\bibinfo {title}
  {Spin-{S}tate-{S}elective {E}xcitation in {S}pin {D}efects of {H}exagonal
  {B}oron {N}itride},}\ }\href@noop {} {\bibfield  {journal} {\bibinfo
  {journal} {Nano Letters}\ }\textbf {\bibinfo {volume} {25}},\ \bibinfo
  {pages} {12067--12074} (\bibinfo {year} {2025})}\BibitemShut {NoStop}%
\bibitem [{\citenamefont {Elko}\ \emph {et~al.}(2024)\citenamefont {Elko},
  \citenamefont {Hassenmayer}, \citenamefont {Higgins}, \citenamefont
  {Lenahan}, \citenamefont {Flatté}, \citenamefont {Fehr}, \citenamefont
  {Craven},\ and\ \citenamefont {Larsen}}]{10.1116/6.0003855}%
  \BibitemOpen
  \bibfield  {author} {\bibinfo {author} {\bibfnamefont {M.~J.}\ \bibnamefont
  {Elko}}, \bibinfo {author} {\bibfnamefont {D.~T.}\ \bibnamefont
  {Hassenmayer}}, \bibinfo {author} {\bibfnamefont {A.~A.}\ \bibnamefont
  {Higgins}}, \bibinfo {author} {\bibfnamefont {P.~M.}\ \bibnamefont
  {Lenahan}}, \bibinfo {author} {\bibfnamefont {M.~E.}\ \bibnamefont
  {Flatté}}, \bibinfo {author} {\bibfnamefont {D.}~\bibnamefont {Fehr}},
  \bibinfo {author} {\bibfnamefont {M.~D.}\ \bibnamefont {Craven}}, \ and\
  \bibinfo {author} {\bibfnamefont {T.~D.}\ \bibnamefont {Larsen}},\ }\bibfield
   {title} {\enquote {\bibinfo {title} {Near zero-field magnetoresistance and
  defects in gallium nitride pn junctions},}\ }\href {\doibase
  10.1116/6.0003855} {\bibfield  {journal} {\bibinfo  {journal} {Journal of
  Vacuum Science \& Technology B}\ }\textbf {\bibinfo {volume} {42}},\ \bibinfo
  {pages} {052205} (\bibinfo {year} {2024})}\BibitemShut {NoStop}%
\bibitem [{\citenamefont {McMillan}, \citenamefont {Harmon},\ and\
  \citenamefont {Flatt\'e}(2020)}]{PhysRevLett.125.257203}%
  \BibitemOpen
  \bibfield  {author} {\bibinfo {author} {\bibfnamefont {S.~R.}\ \bibnamefont
  {McMillan}}, \bibinfo {author} {\bibfnamefont {N.~J.}\ \bibnamefont
  {Harmon}}, \ and\ \bibinfo {author} {\bibfnamefont {M.~E.}\ \bibnamefont
  {Flatt\'e}},\ }\bibfield  {title} {\enquote {\bibinfo {title} {Image of
  {D}ynamic {L}ocal {E}xchange {I}nteractions in the dc {M}agnetoresistance of
  {S}pin-{P}olarized {C}urrent through a {D}opant},}\ }\href {\doibase
  10.1103/PhysRevLett.125.257203} {\bibfield  {journal} {\bibinfo  {journal}
  {Phys. Rev. Lett.}\ }\textbf {\bibinfo {volume} {125}},\ \bibinfo {pages}
  {257203} (\bibinfo {year} {2020})}\BibitemShut {NoStop}%
\end{thebibliography}
\end{document}